\documentclass[12pt]{iopart}
\usepackage{graphicx}
\usepackage{color}
\graphicspath{
{/Users/bertkappen/doc/publ/papers_read/kl_control2017/},
}

\usepackage{siunitx}
\usepackage{url}
\usepackage{subfigure}
\newcommand{\del}[3]{\left#1 #3 \right#2}
\newcommand{\av}[1]{\del{<}{>}{#1}}
\newcommand{\bea}{\begin{eqnarray}}
\newcommand{\eea}{\end{eqnarray}}
\newcommand{\beaa}{\begin{eqnarray*}}
\newcommand{\eeaa}{\end{eqnarray*}}

\newcommand{\bc}{\begin{center}}
\newcommand{\ec}{\end{center}}
\newcommand{\bmp}[1]{\begin{minipage}{#1\textwidth}}
\newcommand{\emp}{\end{minipage}}

\newcommand{\bra}[1]{\del{<}{|}{#1}}
\newcommand{\ket}[1]{\del{|}{>}{#1}}
\newcommand{\cO}{{\cal O}}
\newcommand{\cN}{{\cal N}}
\newcommand{\ben}{\begin{enumerate}}
\newcommand{\een}{\end{enumerate}}

\begin{document}
%\preprint{}
\title{Learning quantum models  from quantum or classical data}
\author{H.J. Kappen}
\address{Donders Institute, Department of Biophysics, Radboud University, the Netherlands}
\ead{b.kappen@science.ru.nl}
\vspace{10pt}
\begin{indented}
%\item[]July 1, 2019
\item\today
\end{indented}

\begin{abstract}
In this paper, we address the problem how to represent a classical data distribution in a quantum system. 
The proposed method is to learn quantum Hamiltonian that is such that its ground state approximates the given classical distribution.
We review previous work on 
the quantum Boltzmann machine (QBM) \cite{kieferova2017tomography,amin2016quantum} and how it can be used to infer quantum Hamiltonians from quantum statistics. We then show how the proposed quantum learning formalism can also be applied to a purely classical data analysis. 
Representing the data as a rank one density matrix introduces quantum statistics for classical data in addition to the classical statistics. 
We show that quantum learning yields results that 
can be significantly more accurate than the classical maximum likelihood approach, both for unsupervised learning and for classification.
The data density matrix and the QBM solution show entanglement, quantified by the quantum mutual information $I$. 
The classical mutual information in the data $I_c\le I/2=C$, with $C$ maximal classical correlations obtained by choosing a suitable orthogonal measurement basis. 
We suggest that the remaining mutual information $Q=I/2$ is obtained by non orthogonal measurements that may violate the Bell inequality. 
The excess mutual information $I-I_c$ may potentially be used to improve the performance of quantum implementations of machine learning or other statistical methods. 

%We illustrate learning a quantum anti-ferromagnetic Heisenberg Hamiltonian and spin glass Hamiltonian from quantum statistics using this approach. The learning problem is convex  and has a unique solution for finite temperature. For zero temperature the problem is ill-posed and the solution is not unique. 

\end{abstract}

\maketitle

\section{Introduction}
Current successes in machine learning \cite{lecun2015deep,mnih2015human,silver2016mastering}
has ignited interesting new connections between machine learning and quantum physics, loosely referred to as quantum machine learning. The transfer of ideas goes in both directions. 
Quantum annealing  \cite{kadowaki1998quantum,heim2015quantum} has been successfully applied to hard optimization problems in machine learning \cite{adachi2015application,benedetti2016estimation}. 
Machine learning methods also find useful applications in quantum physics, such as characterizing the ground state of a quantum Hamiltonian
\cite{carleo2017solving} or to learn different phases of matter \cite{carrasquilla2017machine}.

In addition, there are efforts to exploit quantum mechanical features for learning and coding \cite{biamonte2017quantum}. In particular, attempts have been made to extend probability calculus to quantum density operators
\cite{cerf1997negative,cerf1999quantum,schack2001quantum}. 
Recently, \cite{kieferova2017tomography,amin2016quantum} have proposed learning methods for density matrices called the quantum Boltzmann machine (QBM). 
One approach is to maximize the classical likelihood $L=\sum_s q(s)\log p(s|w)$,
with $p$ the diagonal of the density matrix $\rho$. As the authors remark,  this approach faces difficulties because the gradients of the likelihood are hard to evaluate (see~\ref{appendix:amin}). For this reason, they introduce a lower bound on the likelihood using the Golden-Thomson inequality and maximize this bound. 
But this has the disadvantage that parameters of quantum statistics cannot be learned.  

\cite{kieferova2017tomography} introduce another approach to learning the QBM, which is to minimize the relative entropy between a model density matrix $\rho$ and a target density matrix $\eta$ \footnote{This idea was independently proposed in the arXiv preprint arXiv:1803.11278, which is the draft version of the current article.}.
  This is equivalent to maximizing a quantum likelihood $L=\Tr \eta \log \rho$. 
This approach does not suffer from the above difficulties. The quantum likelihood generalizes maximum likelihood learning to density matrices: In the case that the model density matrix is diagonal, the quantum likelihood reduces to the classical likelihood.

The aim of this paper is to show how the quantum likelihood method can be used to learn classical machine learning problems such as unsupervised and supervised learning. We restrict ourselves to spin systems without latent variables, ie. there is data statistics on all spins. For extensions to latent variable models see \cite{wiebe2019generative}.
For unsupervised learning problems, data can be represented as a classical distribution $q(s)$. 
By generalizing $q$ to a density matrix, the classical statistics (such as expectation values and correlations) are augmented with quantum statistics. 
The quantum statistics provides features of the data that are not available from low order classical statistics and may be useful for learning. 

The paper is organized as follows. 
In section~\ref{learning} we review the unsupervised learning problem and the classical Boltzmann machine. 
In section~\ref{quantum_learning} we review the quantum likelihood and the derivation of QBM learning rule. 
In section~\ref{quantum} we present numerical results that demonstrate how to apply the QBM to quantum tomography, ie. to recover the quantum Hamiltonian or the ground state wave function from measured quantum statistics. We consider the quantum anti-ferromagnetic Heisenberg model and a fully connected quantum spin glass model. 
For finite temperature, this problem is strictly convex and has a unique solution. In this case the Hamiltonian can be fully recovered from the quantum statistics. 
For zero temperature the solution is degenerate. The learned Hamiltonian still reproduces all statistics correctly, but the solution is not unique. For low temperature, the convergence of the learning algorithm slows down dramatically. The results in section~\ref{quantum} confirm previous findings reported in \cite{kieferova2017tomography}.

Sections~\ref{classical} and~\ref{entanglement} contain the main novel material of this paper. In section~\ref{classical} we apply the QBM to learn classical statistical problems, such as unsupervised learning and classification. 
We propose to represent the target data distribution by a rank one density matrix $\eta$, which we call the data state. For unsupervised learning, 
we show that the QBM with only pairwise interactions can learn the parity problem exactly, whereas the BM cannot learn this problem. 
In the case of supervised learning, we learn a density matrix over the joint input-output state space and construct the classifier by conditioning the density matrix on a classical input state. Surprisingly, in this way a large number of hard nonlinear classification problems can be learned that cannot be learned by a classical BM. This approach differs significantly from the recently proposed quantum perceptron \cite{wiersema2019implementing} which learns a density matrix on the output qubit, whose statistics are conditioned on classical input states. 

The data state $\eta$ is a much more complex object than the classical probability distribution $q$ from which it is derived. It displays full quantum features such as non locality and entanglement. In section~\ref{entanglement} we explain what these quantum features mean for classical data.
We can measure $\eta$ in different measurement bases and the classical distribution $q(s)$ corresponds to a particular choice of basis. 
In general, the quantum mutual information $I(\eta)$ between sub systems is larger than
the classical mutual information $I_c(q)$. We conjecture that $I_c(q)\le I(\eta)/2$ in general, which we support by numerical experiments and show that $I_c(q)=I(\eta)/2$ when the sub systems are deterministically related. 
By optimizing the orthogonal measurement basis, the mutual information can be maximized to $I_c(\tilde{q})=I(\eta)/2=C$
where $C$ denotes the maximal classical mutual information. The information is called classical because the statistics of the
measurement outcomes can be described by a classical joint probability distribution $\tilde{q}$ on local variables.
The remaining quantum information $Q=I-C$ resides in non-local features of the state $\eta$.
In section~\ref{bell} 
we illustrate the excess quantum information $Q$ and the violation of the Bell inequality for the simplest possible example of two fully correlated binary variables.

\section{Classical learning\label{learning}}
We first briefly review classical learning. Consider a data set of samples $s^\mu, \mu=1,\ldots N$, where each sample $s^\mu$ is a vector of length $n$. The data set can be written as a so-called empirical probability distribution 
\bea
q(s)=\frac{1}{N}\sum_{\mu=1}^N\delta_{s,s^\mu}\label{qempirical}
\eea
Classical learning can be defined to find a model distribution $p(s)$ that is as close as possible to $q$, where the 'distance' is defined as the relative entropy or Kullback-Leibler divergence between the distributions $q$ and $p$
\bea
KL(q,p)=\sum_s q(s)\log \frac{q(s)}{p(s)}\label{KL}
\eea
Minimizing $KL(q,p)$ with respect to $p$ is equivalent to maximizing the classical likelihood
\bea
L_c=\sum_s q(s)\log p(s)\label{Lclassical}
\eea 

In the case of the classical Boltzmann machine learning problem,
the state space consists of all vectors $s=(s_1,\ldots,s_n)$ with $s_i=\pm 1$ binary spin variables
%Learning can be viewed  as a procedure to find the parameters $w$ 
%of a statistical probability model $p(s|w)$ such that $p$ is as close as possible to $q$. 
%A common notion of distance between distributions  $p$ and $q$ is the Kullback-Leibler divergence or cross-entropy \cite{cover2012elements}
%\bea
%KL(q|p)=\sum_s q(s)\log\frac{q(s)}{p(s|w)}
%\eea
%Minimizing $KL(q|p)$ with respect to $w$ is equivalent to maximizing the likelihood
%\bea
%L_c(w) =\sum_s q(s) \log p(s|w) \label{Lclassical}
%\eea
and $p$ is a Boltzmann distribution 
\bea
p(s|w)=\frac{e^{H(s|w)}}{Z} \qquad Z(w)=\sum_s e^{H(s|w)}\label{p}
\eea
with $H(s|w)=\sum_r H_r(s) w_r$ linear in $w_r$.  
$H_r(s)$ are interaction terms involving typically a small subset of the components of $s$ such for instance $s_i,s_is_j, \ldots$. 
For $p$ of the form Eq.~\ref{p} the likelihood Eq.~\ref{Lclassical} becomes
\bea
L_c(w) =\av{H}_q -\log Z\label{classical_likelihood}
\eea
where $\av{H}_{q}$ is the expectation value of $H$ with respect to the empirical distributions $q$.
The maximization can be performed by gradient ascent on $L_c$: 
\bea
\Delta w_r\propto \frac{\partial L_c}{\partial w_r}= \av{H_r}_q -\av{H_r}_p\label{clearning}
\eea
where we used that $\frac{\partial \log Z}{\partial w_r}=\av{H_r}_p$ and $\av{\ldots}_p$ is expectation with respect to the Boltzmann distribution $p$. Learning stops when the gradients are zero, ie. when the statistics defined by $H_r$ are equal: $\av{H_r}_q=\av{H_r}_p$. 
A common example is when $H(s|w)=\sum_i w_i s_i+\sum_{i>j}w_{ij}s_is_j$, the learning rule becomes
\bea
\Delta w_i\propto  \av{s_i}_q -\av{s_i}_p\qquad \Delta w_{ij}\propto  \av{s_is_j}_q -\av{s_is_j}_p\label{clearning1}
\eea
Eq.~\ref{clearning1} is the well known Boltzmann Machine learning rule \cite{ack85}. 

\section{Quantum learning}
\label{quantum_learning}

In the quantum case, one represents both the data and the model as a density matrix instead of a probability vector. In this section, we review this idea that was first introduced in \cite{kieferova2017tomography}. A density operator $\rho$ is a Hermitian positive semi definite operator. In the case of binary spins, $\rho$ can be represented on an orthogonal basis $\ket{s=(s_1,\ldots,s_n)}$ as a matrix of size $2^n \times 2^n$ with components $\rho(s,s')$. 
We also represent the wave function $\ket{\psi}$ of a quantum system on the same basis by its component vector $\psi(s)=\av{s|\psi}$ (see~\ref{qbm}).
We use this basis dependent representation throughout the paper. 
In section~\ref{entanglement} we discuss measurement in other bases. $\rho$ has trace one $\Tr  \rho =1$ and real eigenvalues $\lambda_s\ge 0$ and $\sum_s \lambda_s=1$. The density matrix is a generalization of a probability distribution and reduces to the latter when it is diagonal. 

The von Neumann, or quantum, entropy of a density matrix $\rho$ is defined as 
\bea
h(\rho)=-\Tr\left(\rho\log\rho\right)\label{quantum_entropy}
\eea
with $\log \rho$ the matrix logarithm of $\rho$ and $\rho\log \rho$ a matrix product.
It is easy to show that $h(\rho)=-\sum_s \lambda_s \log \lambda_s$, with $\lambda_s\ge 0$ the eigenvalues of $\rho$. 
The entropy is maximal when all $\lambda_s$ are equal. The minimal entropy $h(\rho)=0$ when $\lambda_s=\delta_{s,s^*}$ for some state $s^*$. 
In this case $\rho$ is a rank one matrix and can be written as $\rho=\psi\psi^\dagger$ (or $\rho(s,s')=\psi(s)\psi^*(s')$) and is called a pure state. 
When $\rho$ is diagonal: $\rho(s,s')=p(s) \delta_{s,s'}$, the quantum entropy is equal to the classical entropy $h(\rho)=h_c(p)=-\sum_s p(s)\log p(s)$.

The notion of expectation value for probability distributions is generalized for density matrices. The expectation value of a matrix $A$, called an observable, is defined as $\av{A}_\rho=\Tr\left(A \rho\right)$, with $A\rho$ the matrix product.
When $A$ is a Hermitian matrix, $\av{A}_\rho$ is real. 
When $A$ is a diagonal matrix, $\av{A}_\rho=\av{A}_p$ with $p$ the diagonal of $\rho$ and $\av{A}_p$ the classical expectation value. In this case we call $\av{A}_\rho$ a classical statistic (in the basis $\ket{s}$) . 
%For instance when $A(s,s')=s_i\delta_{s,s'}$, $\av{s_i}_\rho$ the mean value of $s_i$ in the distribution $p$. 
When $A$ is a non diagonal matrix, $\av{A}_\rho$ are statistics of $\rho$ that do not have a classical analogue. We call these quantum statistics.

The relative entropy between density matrices $\eta$ and $\rho$ is defined as \cite{carlen2010trace}
\bea
S(\eta,\rho)=\Tr\left(\eta \log \eta\right)-\Tr\left(\eta \log \rho\right)\label{Scross}
\eea
One can show that $S\ge 0$ which follows from Klein's inequality \cite{carlen2010trace}. 
Eq.~\ref{Scross} generalizes the classical relative entropy Eq.~\ref{KL} to density matrices. 
When $\eta$ is the density matrix of the data and $\rho$ is the model density matrix, the quantum learning problem is to find $\rho$ that minimizes $S$. This is equivalent to maximizing the quantum likelihood
\bea
L(\rho)= \Tr \left(\eta \log \rho\right)\label{Lquantum}
\eea

As an immediate generalization of the classical BM case discussed in section~\ref{learning}, we consider model density matrices of the form
\bea
\rho=\frac{1}{Z}e^H\qquad Z=\Tr\left(e^H\right)\label{density_matrix}\qquad H=\sum_r H_r w_r 
\eea
with $H$ the quantum Hamiltonian and $e^{H}$ is the matrix exponential of $H$. 
$H$ and $H_r$ are Hermitian matrices and $w=\{w_r, r=1,\ldots\}$ are real parameters. 
%{\bf In physics, $H$ is the Hamiltonian of the quantum system, and $H_r$ describe the various interaction terms in the Hamiltonian. In classical statistics 
%a model of the form Eq.~\ref{density_matrix} is known as an exponential family model because $H$ depends linearly on the parameters $w_r$. Exponential family models have the advantage that the parameters $w_r$ can be estimated through sufficient statistics \cite{andersen1970sufficiency}.} 
The model Eq.~\ref{density_matrix} is referred to as the quantum Boltzmann machine (QBM) \cite{kieferova2017tomography,amin2016quantum}. 

The quantum likelihood Eq.~\ref{Lquantum} for the QBM  Eq.~\ref{density_matrix} is 
\bea
L(w)=\av{H}_\eta-\log Z\ \label{Lqbm}
\eea

Learning is defined as gradient ascent on the quantum likelihood Eq.~\ref{Lqbm}. 
One can compute $
\frac{\partial}{\partial w_r}  e^H$ 
through the Trotter formula 
$e^H=\lim_{m\to \infty} \left(e^{H/m}\right)^m$:
\beaa
\frac{\partial}{\partial w_r} e^H&=& \lim_{m\to\infty}\frac{H_r}{m}\underbrace{e^{H/m}\ldots e^{H/m}}_{m~\mathrm{terms}}+\ldots \\
&+& \underbrace{e^{H/m}\ldots e^{H/m}}_{a~\mathrm{terms}}\frac{H_r}{m} \underbrace{e^{H/m}\ldots e^{H/m}}_{m-a~\mathrm{terms}}+\ldots + \ldots \underbrace{e^{H/m}\ldots e^{H/m}}_{m~\mathrm{terms}}\frac{H_r}{m}\\
&=&\int_0^1 dt e^{Ht}H_r e^{H(1-t)}
\eeaa
Thus $\frac{\partial}{\partial w_r}\Tr \left(e^H\right)=\Tr \left(H_r e^H\right)$ and $
\frac{\partial}{\partial w_r} \log Z\av{H_r}_\rho$. 
Since $\av{H}_\eta =\sum_r w_r \av{H_r}_\eta$ we get 
\beaa
\Delta w_r\propto \frac{\partial}{\partial w_r} L =\av{H_r}_\eta -\av{H_r}_\rho \label{qlearning}
\eeaa

For the rest of the paper we consider binary quantum spin systems with Hamiltonian
\bea
H=\sum_{i=1}^n\sum_{k=x,y,z} w_i^k \sigma_i^k+\sum_{i=1,j>i}^n \sum_{k=x,y,z}  w_{ij}^{k} \sigma_i^k\sigma_j^k\label{Hamiltonian}
\eea
$\sigma_i^{x,y,z}$ are Pauli spin 1/2 operators (see \ref{qbm}). 
For this Hamiltonian, the learning rule Eq.~\ref{qlearning} becomes
\bea
\Delta w_i^k &=& \epsilon\left(\av{\sigma_i^k}_\eta-\av{\sigma_i^k}_\rho\right)\qquad
\Delta w_{ij}^k = \epsilon\left(\av{\sigma_i^k\sigma_j^k}_\eta-\av{\sigma_i^k\sigma_j^k}_\rho\right) 
\label{qlearning1}
\eea
with $k=x,y,z$ and $\epsilon>0$ the learning rate. 
The QBM reduces to the classical BM when $k$ takes only the value $k=z$ in Eqs.~\ref{Hamiltonian} and~\ref{qlearning1}.

%{\color{blue} (Begin moved from appendix B)
%The learning algorithm requires computation of the statistics $\av{\ldots}_\rho$ given $H$ for each learning iteration. Since the target $\eta$ is a rank one density matrix, we often find that the solution $\rho$ is of low rank. 
%We can therefore accelerate the computation of these statistics by making a low rank approximation of $\rho$. 
%(End moved)}

\section{Learning a quantum Hamiltonian}
\label{quantum}
\begin{figure}
\bc
\subfigure[
Hamiltonian couplings and pair wise statistics.]
{\includegraphics[width = 0.5\textwidth]{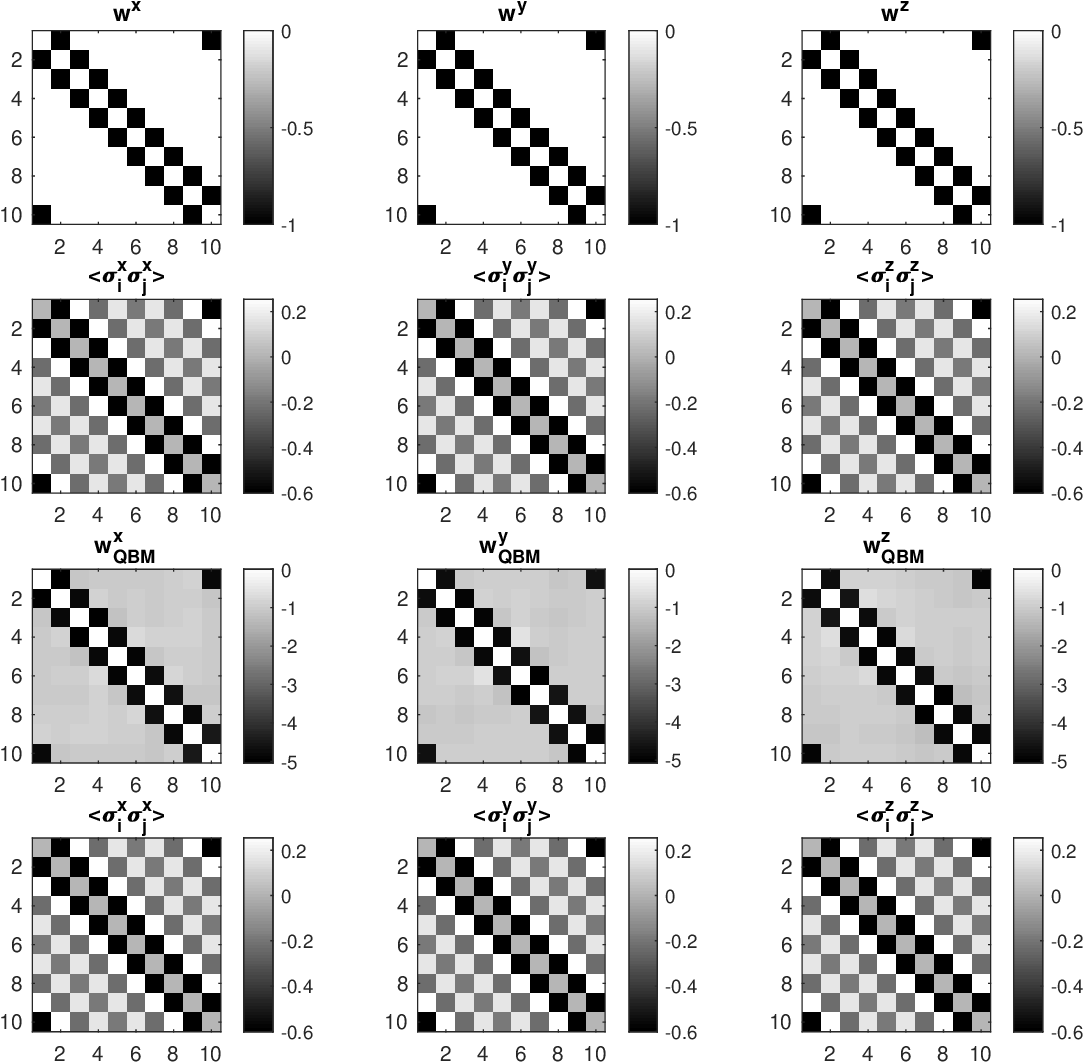}}\\
\subfigure[
Relative entropy versus learning iteration.
%(d) Error in the couplings $w^{x,y,z}_{ij}, w^{x,y,z}_{i}$ versus learning iteration for $\beta=1$ (blue), $\beta=2$ (black) and $\beta=\infty$ (red). 
]
{\includegraphics[width = 0.4\textwidth]{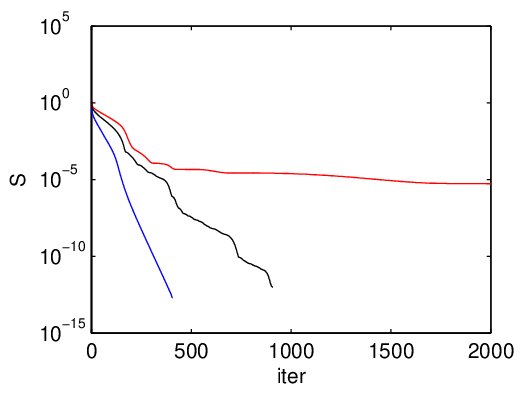}}
\hspace*{1cm}
\subfigure[
Ground state wave function $\phi$ of the learned QBM and classical BM distribution.]{\includegraphics[width = 0.4\textwidth]{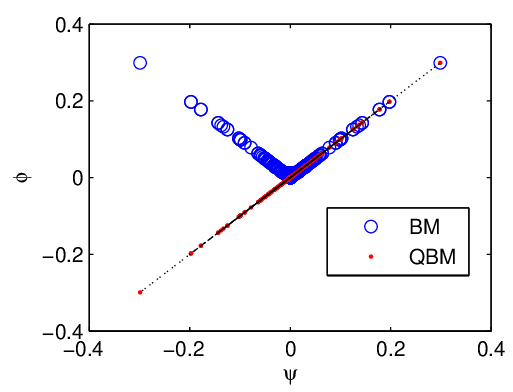}}
\ec
\caption{(Color online) Quantum learning of the one-dimensional anti ferromagnetic Heisenberg chain of $n=10$ spins with periodic boundary conditions.
(a) Top row: The Hamiltonian Eq.~\ref{Hamiltonian} has couplings $w^{x,y,z}_{ij}=-1$ for nearest neighbors and $w^{x,y,z}_{ij}=0$ otherwise and external fields 
$w_i^{x,y,z}=0$.  
Second row: The quantum statistics $\av{\sigma_{i}^k\sigma_j^k}_{\eta_\beta}, k=x,y,z$ for $\beta=\infty$. 
Third row: The QBM couplings $w^{x,y,z}_{ij}$ after converged learning from the $\beta=\infty$ statistics.
Fourth row: The QBM statistics $\av{\sigma_{i}^k\sigma_j^k}_\rho, k=x,y,z$. 
(b) Relative entropy $S(\eta,\rho)$ versus learning iteration for $\beta=1$ (blue), $\beta=2$ (black) and $\beta=\infty$ (red). 
%(d) Error in the couplings $w^{x,y,z}_{ij}, w^{x,y,z}_{i}$ versus learning iteration for $\beta=1$ (blue), $\beta=2$ (black) and $\beta=\infty$ (red). 
(c) Scatter plot of the $2^n$ components of the ground state wave function $\phi(s)$ of the learned quantum Hamiltonian versus 
the components of the true ground state wave function $\psi(s)$ (red). For comparison $\sqrt{p_\mathrm{bm}(s)}$ versus $\psi(s)$ with $p_\mathrm{bm}$ the solution of the classical BM (blue). 
}
\label{file4_afh}
\end{figure}

In this section we provide some examples of learning a density matrix of an unknown quantum system from observed quantum statistics. The results in this section confirm previous findings reported in \cite{kieferova2017tomography}.
As a first example, we consider the anti-ferromagnetic Heisenberg model  in 1 dimension with true Hamiltonian $H$ given by Eq.~\ref{Hamiltonian} with couplings $w^{x,y,z}_{ij}=-1$ for nearest neighbors and $w^{x,y,z}_{ij}=0$ otherwise  (Figure~\ref{file4_afh}a, top row) and external fields 
$w_i^{x,y,z}=0$. 
From $H$, the data density matrix $\eta_\beta=\frac{1}{Z}e^{\beta H}$ is constructed for $\beta=1,2,\infty$ and the quantum statistics
$\av{\sigma_i^{k}}_{\eta_\beta}=0$ and $\av{\sigma_{i}^k\sigma_j^k}_{\eta_\beta}$  (shown for $\beta=\infty$ in Figure~\ref{file4_afh}a, second row) are computed. 
These statistics are used to train the QBM using the learning rule Eq.~\ref{qlearning1} that minimizes the relative entropy $S(\eta,\rho)$ between the data density matrix $\eta$ and the model density matrix $\rho$ (Figure~\ref{file4_afh}b).  Learning stops when the change in $S$ approaches machine precision or when $2000$ iterations are reached. 
For $\beta=1,2$, learning converges fast to the optimal solution ($S(\eta_{\beta=1},\rho)=\num{2e-13}$ and $S(\eta_{\beta=2},\rho)=\num{1e-12}$) and the Hamiltonian parameters are accurately reconstructed (RMS error is $\num{5e-7}$ and $\num{2e-6}$, respectively). 
For all $\beta$, the learned density matrix accurately models the quantum statistics since this is the learning fixed point in Eq.~\ref{qlearning1} (for $\beta=\infty$ compare Figure~\ref{file4_afh}a, second and fourth row).

For lower temperature, learning becomes increasingly difficult. 
For the $\beta=\infty$ learning problem, $\eta_{\beta=\infty}=\psi\psi^\dagger$ is a  rank one density matrix and $\rho\propto e^{\sum_r w_r H_r} \to \phi\phi^\dagger$ slowly approaches a rank one density matrix as the couplings $w_r$ diverge during learning ($S(\eta_{\beta=\infty},\rho)=\num{5e-6}$ after 2000 iterations). 
Figure~\ref{file4_afh}c shows a scatter plot of $\phi$ versus $\psi$ which has the same small error. 
However, the Hamiltonian parameters are inaccurately reconstructed.
In order to compare the learned parameters with their true values, we estimate an effective $\beta_\mathrm{eff}$ from the learned parameters and define the RMS error as $\min_{\beta_\text{eff}} \sum_r (\beta_\mathrm{eff} w^\mathrm{true}_r-w_r)^2$. We obtain  $\beta_\mathrm{eff}=5.6$ and the RMS error is $0.18$. 
This error is large, but still sufficiently small to detect the zero and nonzero couplings (compare Figure~\ref{file4_afh}a, first and third row). 

%\begin{figure}
%\bc
%\includegraphics[width=0.3\textwidth]{file4AFHc.eps}
%\includegraphics[width=0.3\textwidth]{print_fig1c1.eps}
%\includegraphics[width=0.3\textwidth]{file4AFHpsi.eps}
%%\includegraphics[width=0.3\textwidth]{print_fig1c2.eps}
%%\includegraphics[width=0.3\textwidth]{print_fig1c3.eps}
%\ec
%\caption{(Color online) Quantum learning of the $1$ dimensional anti-ferromagnetic Heisenberg chain of $n=10$ spins with periodic boundary conditions. 
%(a) Top row: The Hamiltonian Eq.~\ref{Hamiltonian} has couplings $w^{x,y,z}_{ij}=-1$ for nearest neighbours and $w^{x,y,z}_{ij}=0$ otherwise and external fields 
%$w_i^{x,y,z}=0$.  
%Second row: The quantum statistics $\av{\sigma_{i}^k\sigma_j^k}_\eta$ for $\beta=\infty$. 
%Third row: Reconstructed couplings $w^{x,y,z}_{ij}$ for $\beta=\infty$.
%Fourth row: Reconstructed quantum statistics $\av{\sigma_{i}^k\sigma_j^k}_\rho$ for $\beta=\infty$.
%(b)  entropy $S(\eta,\rho)$ versus learning iteration for $\beta=1$ (blue), $\beta=2$ (black) and $\beta=\infty$ (red). 
%%(d) Error in the couplings $w^{x,y,z}_{ij}, w^{x,y,z}_{i}$ versus learning iteration for $\beta=1$ (blue), $\beta=2$ (black) and $\beta=\infty$ (red). 
%(c) Scatter plot of the $2^n$ components of the ground state wave function $\phi$ of the learned quantum Hamiltonian versus the components of the true ground state wave function $\psi$ (red). For comparison $\sqrt{p_\mathrm{bm}}$ versus $\psi$ with $p_\mathrm{bm}$ the solution of the classical BM (blue).  
%}
%\label{file4_afh}
%\end{figure}

We also train a classical BM on these same problems using the learning rule Eq.~\ref{clearning1}.
The BM does not yield a good solution for any $\beta$: $S(\eta_{\beta=1},\rho_\mathrm{bm})=3.08, S(\eta_{\beta=2},\rho_\mathrm{bm})=3.73, S(\eta_{\beta=\infty},\rho_\mathrm{bm})=4.01$ with $\rho_\mathrm{bm}=\mathrm{diag}(p_\mathrm{bm})$ and $p_\mathrm{bm}$ the learned BM distribution. 
The BM correctly reproduces the classical statistics $w_{ij}^z,w_i^z$ for which it is optimized, but not the quantum statistics.
Surprisingly for $\beta=\infty$, $p_\mathrm{bm}$ correctly reproduces the absolute values  $\sqrt{p_\mathrm{bm}(s)}\approx |\psi(s)|$ for all states $s$, but of course not the correct signs (Figure~\ref{file4_afh}c).

%\begin{figure}
%\bc
%\includegraphics[width=0.3\textwidth]{file4AFHpsi.eps}
%\ec
%\caption{(Color online) Quantum learning of the $1$ dimensional anti-ferromagnetic Heisenberg chain of $10$ spins with periodic boundary conditions. 
%The Hamiltonian Eq.~\ref{Hamiltonian} has couplings $w^{x,y,z}_{ij}=-1$ for nearest neighbours and $w^{x,y,z}_{ij}=0$ otherwise
%($w^{x,y,z}$  top row SM Figure~1) and external fields 
%$w_i^{x,y,z}=0$ . 
%From $H$, the ground state wave function $\psi$ and the quantum statistics
%$\av{\sigma_i^{k}}_\eta=0$ and $\av{\sigma_{i}^k\sigma_j^k}_\eta$ are computed (second row SM
%Figure~1). 
%All statistics are used to learn the quantum Boltzmann machine (Eq.~\ref{qlearning1}) and all
%classical statistics are used to learn the classical Boltzmann machine (SM Eq.~4).
%Scatter plot of the true ground state wave function $\psi$ versus the ground state wave function of the learned Hamiltonian. The QBM correctly reproduces $\psi$. For comparison $\sqrt{p_\mathrm{bm}(s)}$, with $p_\mathrm{bm}(s)$ the solution of the classical BM, is also plotted, which correctly reproduces $|\psi|$ but not the sign.   
%The learned quantum couplings $w_\mathrm{QBM}^{x,y,z}$ (third row left SM Figure~1) resemble the
%original couplings $w^{x,y,z}$ up to an overall scaling and perfectly reproduce the quantum
%statistics (fourth row left SM Figure~1).
%The learned classical couplings $w_\mathrm{BM}$ (third row right SM Figure~1) do not resemble the
%original couplings $w^{x,y,z}$ and only reproduce the classical statistics (fourth row right SM
%Figure~1).
%}
%\label{file4_afh}
%\end{figure}

\begin{figure}
\bc
\subfigure[
Relative entropy.]
{\includegraphics[width = 0.3\textwidth]{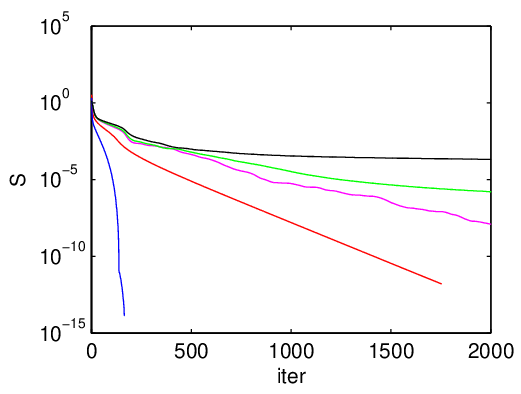}}
\subfigure[RMS parameter error.]
{\includegraphics[width = 0.3\textwidth]{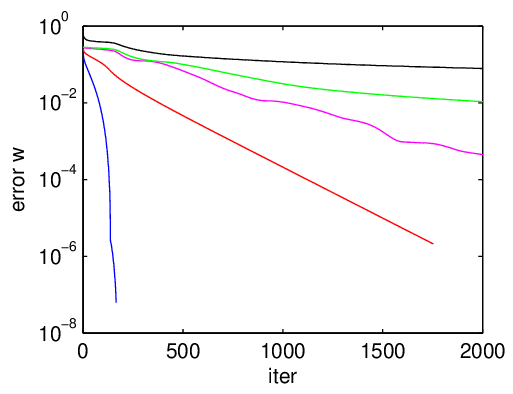}}
\ec
\caption{(Color online) Quantum learning of fully connected quantum spin glass Hamiltonian of $n=10$ spins with random couplings. 
(a) Relative entropy $S(\eta,\rho)$ versus learning iteration for $\beta=0.5$ (blue), $\beta=1$ (red), $\beta=3$ (magenta), $\beta=5$ (green) and $\beta=\infty$ (black). 
(b) RMS error of the estimated Hamiltonian parameters $w_i^{x,z},w_{ij}^{x,y,z}$ versus learning iteration for these same $\beta$ values. 
%(b) Scatter plot of the ground state wave function $\phi$ of the learned quantum Hamiltonian versus the true ground state wave function $\psi$ (red). For comparison $\sqrt{p_\mathrm{bm}(s)}$ versus $\av{s|\psi}$ with $p_\mathrm{bm}(s)$ the solution of the classical BM (blue).  
%(c) Scatter plot for $\beta=1$ learning problem of estimated Hamiltonian parameters $w_i^{x,z},w_{ij}^{x,y,z}$ rescaled by $\beta_\mathrm{eff}=6.4$ versus the true values. 
}
\label{file4_sk}
\end{figure}

As a second example we repeat the above experiment for a fully connected quantum spin glass.  The true Hamiltonian $H$ (Eq.~\ref{Hamiltonian}) on $n=10$ spins has random
couplings $w_{ij}^k\sim \cN\left(0,\frac{1}{\sqrt{n}}\right)$ and random external fields $w_i^{x,z}\sim  \cN\left(0,1\right)$, with $\cN(\mu,\sigma)$ is the Gaussian distribution with mean $\mu$ and standard deviation $\sigma$.
The results are shown in Figure~\ref{file4_sk} and are qualitatively similar to the AFH model in Figure~\ref{file4_afh}. For small $\beta$, learning converges fast to the optimal solution and convergence is slower for increasing $\beta$ both in terms of $S$ (Figure~\ref{file4_sk}a) and the RMS error in the learned Hamiltonian parameters (Figure~\ref{file4_sk}b). 
For $\beta=\infty$, convergence is very slow ($S(\eta_{\beta=\infty},\rho)=\num{2.1e-4}$ after 2000 iterations) and the RMS error in the learned Hamiltonian parameters is $0.0785$ ($\beta_\mathrm{est}=6.4$).  

Note, that all these problems have a solution $\rho=\eta_\beta$ with $S=0$ for any $\beta$, but the QBM learning rule has increasing difficulty to find this solution for large $\beta$. The reason is the following. 
When $\eta>0$ (its eigenvalues are positive), then $S$ is strictly convex in $\rho$ and $S=0$ if and only if $\rho=\eta$ (see \cite{carlen2010trace}). In this case there is a unique optimum and the parameters of the target Hamiltonian can be recovered exactly from the quantum statistics. Examples are the finite temperature quantum Hamiltonian systems discussed above. Instead, for $\beta=\infty$, the density matrix $\eta_{\beta=\infty}=\psi\psi^\dagger$ is not positive (it has eigenvalues zero), $S$ is not strictly convex \cite{carlen2010trace} and the reconstructed Hamiltonian is not unique. 
However, provided that the QBM model Hamiltonian contains all relevant interaction terms, the QBM can accurately reconstruct the density matrix from  the observed statistics. Therefore it can also accurately represent any (non local) quantum feature of such system,  such as entanglement. 

Quantum state tomography as discussed in this section cannot be achieved by optimizing a classical likelihood. However,  \cite{torlai2017many} show that by using multiple bases representations of the wave function simultaneously, one can accurately reconstruct the ground state wave function using a maximum likelihood approach. They do not reconstruct the Hamiltonian. 

It should be noted that the data statistics that are required for learning are prescribed by the model Hamiltonian that is assumed. Each term $H_r$ in Eq.~\ref{density_matrix} introduces a parameter $w_r$ that requires statistics $\av{H_r}$ for learning. This has the advantage that the accuracy and complexity of the learning problem can be adjusted by changing the Hamiltonian. This is in contrast with \cite{ferrie2014self,granade2017practical} who present methods that can be used to reconstruct a pure state \cite{ferrie2014self} or mixed state \cite{granade2017practical} without assuming a particular model (such as the exponential represention in terms of a Hamiltonian Eq.~\ref{density_matrix}). Such 'model free' approach clearly requires much more data for learning and does not reconstruct the Hamiltonian of the system.

\section{Quantum statistics in classical data}
\label{classical}
We can also apply the QBM to learn classical statistical learning problems. We first consider unsupervised learning.
In this case, the data is represented as a classical distribution Eq.~\ref{qempirical}. We define a rank one density matrix 
\bea
\eta=\ket{\psi}\bra{\psi} \qquad \psi(s) =\av{s|\psi}=\sqrt{q(s)}e^{i\alpha(s)}\label{eta_classical}
\eea
which we refer to as the data state. $\alpha(s)$ is an arbitrary $s$ dependent phase. 
The definition Eq.~\ref{eta_classical} implies classical and quantum statistics for the classical data distribution $q$. 
Note, that the expectation of an observable $A$ is
\bea
\av{A}_\eta=\sum_{s,s'}A(s,s') \sqrt{q(s)q(s')}e^{i\alpha(s)-i\alpha(s')}\label{avA}
\eea
For classical statistics, $A(s,s')=A(s)\delta_{s,s'}$ is diagonal and $\av{A}_\eta =\sum_s A(s) q(s)$ is equal to the classical expectation value. 
When $A$ is not diagonal, $\av{A}_\eta$ defines new 'quantum' statistics for $q$.  
Thus, the quantum statistics defined in Eq.~\ref{avA} generalize the classical statistics. Note, that the classical statistics are linear in $q$ and the quantum statistics are quadratic in $\sqrt{q}$. 

We propose the data state $\eta$ as the target density matrix for the quantum likelihood Eq.~\ref{Lquantum}. The optimal solution $\rho$ is a density matrix that represents the quantum statistics in the classical data and has no equivalent in terms of a probability distribution. 
When $\rho$ is diagonal, $\rho(s,s')=p(s)\delta_{s,s'}$,  the quantum likelihood Eq.~\ref{Lquantum} reduces to the classical likelihood Eq.~\ref{Lclassical}.
Thus, quantum learning is the generalization of classical learning to density matrices. 
It can find better solutions because the optimization is over a larger class of models. In addition, it learns from quantum statistics that have no classical analogue.

The quantum likelihood depends on the phase $\alpha(s)$ in Eq.~\ref{eta_classical} which can take infinitely many values. Therefore, we need to define $\alpha(s)$ in order to properly define the learning objective. However, we can absorb $\alpha(s)$ in the model $\rho$ in the following way. 
Denote $\eta_0$ as the density matrix with $\alpha(s)=0$. Then Eq.~\ref{eta_classical} states that $\eta=U \eta_0 U^\dagger$ with
$U(s',s)=\delta_{s,s'} e^{i\alpha(s)}$ a diagonal unitary matrix. 
Then 
\beaa
L=\Tr\left(\eta \log \rho\right) = \Tr\left(\eta_0 \log( U^\dagger \rho U)\right)
\eeaa
Thus, a non-zero $\alpha(s)$ is equivalent to a change in model density matrix from $\rho$ to $U^\dagger \rho U$. 
The phases $\alpha$ can in principle be learned together with the other parameters that define $\rho$. The resulting learning rules are similar (but different) from Eq.~\ref{qlearning1} and their evaluation is of similar computational complexity. 

We will not consider this generalization here and choose $\alpha(s)=0$ for the numerical experiments in this section. We will leave the issue of optimizing $\alpha$ as a topic for future research. 
The data statistics in the QBM learning rule Eq.~\ref{qlearning1} are then
\bea
\bmp{0.29}
\centering
$\begin{array}{r@{{}\mathrel{=}{}}l}
\av{\sigma_i^x}_\eta & \sum_s\sqrt{q(F_is)q(s)}\\[\jot]
\av{\sigma_i^y}_\eta &0 \\[\jot]
\av{\sigma_i^z}_\eta &\sum_s s_i q(s)
\end{array}$
\emp
\bmp{0.29}
\centering
$\begin{array}{r@{{}\mathrel{=}{}}l}
\qquad \av{\sigma_i^x\sigma_j^x}_\eta & \sum_s \sqrt{q(F_iF_js)q(s)}\\[\jot]
\qquad \av{\sigma_i^y\sigma_j^y}_\eta &-\sum_s s_is_j \sqrt{q(F_iF_js)q(s)}\\[\jot]
\qquad \av{\sigma_i^z\sigma_j^z}_\eta & \sum_s s_i s_j q(s)
\end{array}$
\emp
\label{spin_statistics}
\eea
Since $\eta$ is real symmetric, the expectation of complex Hermitian observables such as $\av{\sigma_i^y}_\eta$ are zero.
In general, computation of these data statistics requires $2^n$ operations.  However, when $q$ is given in terms of a data set of $N$ patterns, as in Eq.~\ref{qempirical}, the classical statistics can be computed linear in $N$. For the quantum statistics we 
compute the set of uniquely occurring patterns $\{s^a, a=1,\ldots N_\mathrm{unique}\}$ in the data set with $N_\mathrm{unique}\le N$ and the number of times $N(s^a)>0$ that each of these patterns occurs . This computation requires $\cO(N\log N)$ operations. Subsequently, we can compute each quantum statistics quadratic in $N_\mathrm{unique}$\footnote{For instance, for the computation of $\av{\sigma_i^x}=\frac{1}{N}\sum_{a=1}^{N_\mathrm{unique}}\sqrt{N(s^a)N(F_i s^a)}$. For each $s^a$, we compute $N(F_i s^a)$ by scanning the list of all nonzero $N(s^a)$ for $N(F_i s^a)$. This requires less than $N_\mathrm{unique}$ steps. Possibly, this computation can be accelerated.

%=\frac{1}{N}\sum_s N_s \sqrt{\frac{N_{F_is}}{N_s}}=\frac{1}{N}\sum_\mu \sqrt{\frac{N_{F_is^\mu}}{N_{s^\mu}}}
%=\frac{1}{N}\sum_{\mu, \mathrm{unique}}\sqrt{N_{s^\mu}N_{F_i s^\mu}}=b_i^x\\

}

The QBM can learn classical data problems problems significantly more accurate than the BM, for two reasons. The obvious reason is that the QBM has about 3 times as many parameters than the BM when both include first and second order interactions. Therefore, the QBM should fit the data always at least as good as the BM. However, this does not necessary generalize to new data. 
The second reason is that the QBM not only has more parameters, but these parameters are constrained by additional statistics (the quantum statistics). So the QBM sees more information about the distribution $q$ than the BM. This also improves learning. We illustrate these ideas with the following two numerical examples.
 
\begin{figure}
\bc
\subfigure[
One repeat of neural activity of 160 salamander retinal ganglion cells.]
{\includegraphics[width = 0.9\textwidth]{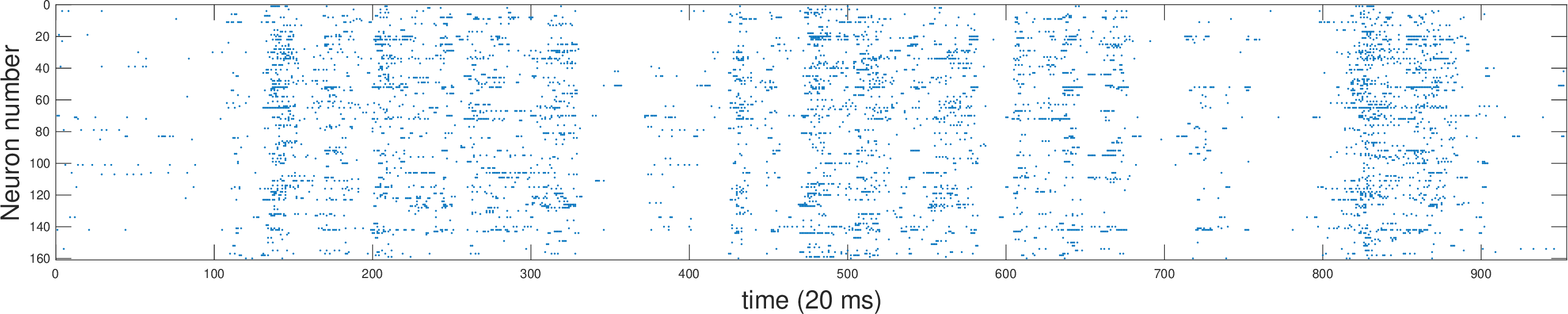}\label{salamander_data}}\\
\subfigure[
$p_\mathrm{bm}$ and $p_\mathrm{qbm}$ versus $q$.]
{\includegraphics[width = 0.4\textwidth]{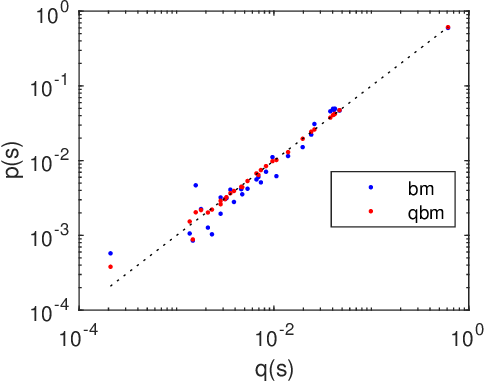}\label{salamander_scatter}}
\subfigure[
Histogram of KL divergences of BM and QBM on 28 independent test sets.]
{\includegraphics[width = 0.4\textwidth]{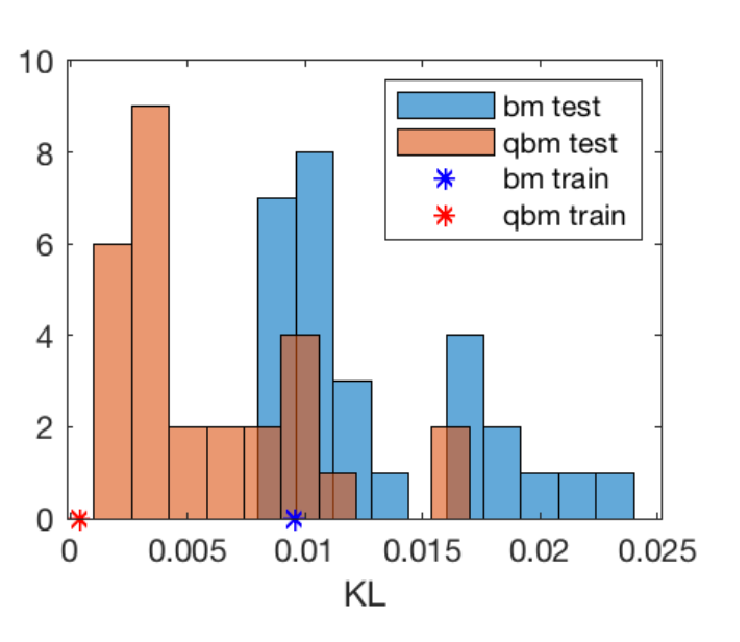}\label{salamander_histo}}
\ec
\caption{(Color online). QBM and BM learning on neural spike data. 
Data consists of 297 repeats of 19 second simultaneous recordings of 160 neurons. Time is divided into 20 ms time bins. The data is binary and indicates whether a given cell elicited at least one spike in a given bin.  One repeats is shown in (a). 
(b) Scatter plot of BM solution $p_\mathrm{bm}$ and rank one approximation $p_\mathrm{qbm}$ of QBM (see text) versus the training data distribution $q$. 
(c) Histogram of KL divergences of BM and QBM on 28 independent test sets. KL divergence on the training set is indicated by *. 
}
\label{salamander}
\end{figure}

The first data set was collected from preprocessed multi-electrode array recording from 160 salamander retinal ganglion cells responding to 297 repeats of a 19 s natural movie
\cite{tkavcik2014searching}\footnote{See \url{https://datarep.app.ist.ac.at/61/2/bint_fishmovie32_100.zip} for further details and the data.}. See Figure~\ref{salamander_data} for data from one repeat.
We selected the 5 neurons with highest average firing rates. 
We used data from 10 repeats to train the BM and the QBM. The QBM solution is significantly better than the  BM solution ($S(\eta,\rho_\mathrm{bm})= 1.79$ and $S(\eta,\rho_\mathrm{qbm})= \num{1.06e-2}$ with $\rho_\mathrm{bm}=\mathrm{diag}(p_\mathrm{bm})$ and $p_\mathrm{bm}$ the learned BM distribution), indicating that the QBM represents the (diagonal and non diagonal) statistics of $\eta$ better. 

We can also assess how well the QBM and BM capture the classical statistics. 
$\rho_\mathrm{qbm}$ is close to rank one: its largest eigenvalue is $\num{0.9987}$. $\rho_\mathrm{qbm}$ is thus well approximated by its rank one approximation: $\rho_\mathrm{qbm}\approx \psi\psi^\dagger$ with $\psi$ the extreme eigenvector of $\rho_\mathrm{qbm}$. From Eq.~\ref{eta_classical}, this implies that $\rho_\mathrm{qbm}$ approximately represents the classical distribution $p_\mathrm{qbm}(s)=\left|\psi(s)\right|^2$.
The $KL$ divergences are $KL(q|p_\mathrm{bm})=\num{9.57e-3}$ and $KL(q|p_\mathrm{qbm})=\num{4.65e-4}$, which shows that the QBM also represents the classical statistics better than the BM.  A scatter plot of $p_\mathrm{bm}$ and $p_\mathrm{qbm}$ versus the original data distribution $q$ is shown in Figure~\ref{salamander_scatter}. We tested the generalization performance of the BM and QBM solution on 28 test sets, each consisting of 10 repeats, by computing the KL divergences $KL(q_i|p_\mathrm{(q)bm})$ between the (Q)BM solution and these 28 empirical distributions $q_i$. A histogram of these $KL$ divergences are shown in Figure~\ref{salamander_histo}. We also compute the relative entropies $S(\eta_i|\rho_\mathrm{(q)bm})$ with $\eta_i$ the rank one density matrix from $q_i$ and find $S(\eta_i|\rho_\mathrm{bm})=1.81\pm 0.02$ and $S(\eta_i|\rho_\mathrm{qbm})=0.120\pm 0.057$. The conclusion is that the QBM solution is significantly more accurate than the BM, both in terms of classical and quantum statistics, and this accuracy generalizes to unseen data. 

\begin{figure}
\bc
\includegraphics[width=0.45\textwidth]{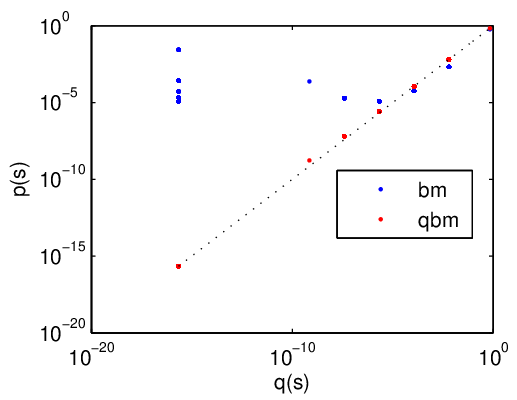}
\ec
\caption{(Color online). Scatter plot of $p_\mathrm{bm}(s)$ and $p_\mathrm{qbm}(s)$ versus the true probability $q(s)$ for the biased parity problem ($\theta=1$) on $n=10$ spins. }
\label{parity_bm}
\end{figure}

%This was tested on a set of random instances proposed in \cite{amin2016quantum} (see SM Figure~2). 
%Average results over 10 instances give 
%quantum cross entropies $S(\eta,\rho_\mathrm{qbm})=0.282\pm 0.14$ and $S(\eta,\rho_\mathrm{bm})=3.53\pm 0.11$.
The second data set shows that the QBM can learn problems that cannot be learned by the classical BM. As an extreme example,  we consider the parity problem on $n$ spins.
The data distribution is $q(s)\propto q_0(s)\exp(\theta \sum_i s_i)$, where $q_0(s)=1,0$ when the parity of $s$, ie. $\prod_i s_i$, is even or odd, respectively and $\theta$ biases the spins to a nonzero mean value. This problem cannot be learned by a classical BM, because it can only effectively model up to second order classical statistics and the parity problem requires knowledge of $n$th order classical statistics. The QBM can learn this problem perfectly. For $\theta=0$, the optimal solution $\rho_\mathrm{qbm}=\frac{1}{2}\left(\psi_1\psi_1^\dagger + \psi_2\psi_2^\dagger\right)$ is a rank two density matrix with $\psi_{1,2}$ orthonormal vectors that are linear combinations of the even parity ($\sqrt{q_0}$) and odd parity ($\sqrt{1-q_0}$) solutions. The reason is that the data statistics $\av{\ldots}_\eta$ are identical for the even and odd parity problem. 
The $\theta$ dependent term biases the solution to even ($\theta>0$) or odd ($\theta<0$) parity. For $\theta=1$ and $n=10$ the solution $\rho_\mathrm{qbm}$ is close to a rank one (largest eigenvalue is $0.9973$) and $S(\eta,\rho_\mathrm{qbm})=\num{4.69e-03}$. The BM solution has $S(\eta,\rho_\mathrm{bm})=2.37$. 
As above, we approximate $\rho_\mathrm{qbm}\approx \psi\psi^\dagger$ with $\psi$ the extreme eigenvector of $\rho_\mathrm{qbm}$ and define $p_\mathrm{qbm}(s)=\left|\psi(s)\right|^2$. The KL divergences are $KL(q|p_\mathrm{bm})=\num{0.451}$ and $KL(q|p_\mathrm{qbm})=\num{1.31e-5}$. 
In Figure~\ref{parity_bm} we show that $p_\mathrm{qbm}$ correctly represents the biased parity problem while $p_\mathrm{bm}$ does not capture any structure in the data.

\subsection{Classification}
We can use the QBM for binary classification. Suppose that input vectors are $x=s_{1:n-1}$ and the class labels are $y=s_n=\pm 1$. Denote  $y=f(x)$ a deterministic mapping that maps each input pattern $x$ onto a class label $y$. We define the data distribution as 
$q(s)=q(x)q(y|x)$ with $q(y|x)=\delta_{y,f(x)}$ to train the BM and QBM. 

For the QBM we define the conditional (classical) probability of $y$ given $x$ as
\beaa
p_\mathrm{qbm}(y|x)=\frac{|\psi(x,y)|^2}{\sum_y |\psi(x,y)|^2}
\eeaa
with $\psi$ the ground state wave function of the learned Hamiltonian.
\footnote{
Alternatively, we could have defined a classifier based on the entire density matrix $\rho$ as
\beaa
p(y|x)=\frac{\rho_\mathrm{qbm}((x,y),(x,y))}{\sum_y \rho_\mathrm{qbm}((x,y),(x,y))}
\eeaa
However, numerically we found that using $\rho_\mathrm{qbm}$ rather than the ground state wave function does not improve classification. The reason may be that successful classification requires a solution $\rho\approx \eta$, which is rank one.}

Similarly, for the BM we define
\beaa
p_\mathrm{bm}(y|x)=\frac{p_\mathrm{bm}(x,y)}{\sum_y p_\mathrm{bm}(x,y)}
\eeaa
Since the BM has pairwise interactions, $p_\mathrm{bm}(y|x)$ can be written as $p_\mathrm{bm}(y|x)=\sigma\left(y\sum_i w_i x_i\right)$ with $\sigma(x)$ is a sigmoid function. 
Therefore, the BM yields a linear classifier, and can thus only correctly classify linearly separable problems. 
For the QBM this is not true and $p_\mathrm{qbm}(y|x)$ can also correctly classify nonlinear classification problems.
For both models, we define the classifier by selecting the most likely class label: $y(x)=\mathrm{argmax}_{y=\pm 1} \  p_\text{(q)bm}(y|x)$.

\begin{figure}
\bc
\includegraphics[width=0.4\textwidth]{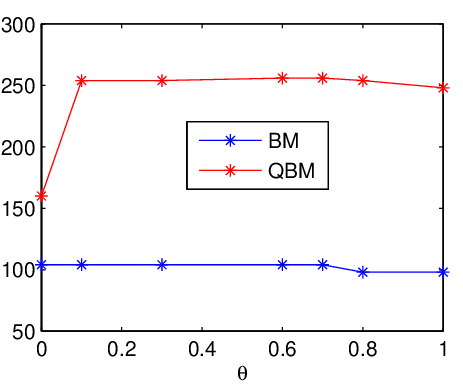}
% generated with file24.m
\ec
\caption{Number of correctly classified problems for the BM and QBM as a function of $\theta$. For the BM, 104 problems are correctly classified for $0\le \theta\le 0.7$. 
For the QBM, 254 or more problems are correctly classified for $0.1\le \theta\le 0.8$ and all 256 problems are correctly classified for $\theta\approx$ 0.6--0.7. 
}
\label{theta_dependence}
\end{figure}

%We have already seen that the QBM can perfectly learn the biased parity problem. Therefore, the QBM is a biased parity detector with zero classification error, whereas the BM will make many classification error. 
To assess the quality of the QBM as a classifier we  use $n=4$ and consider all $2^{2^{n-1}}=256$ binary functions on $3$ inputs. 
We train the BM and QBM on all these problems with $q(x)\propto \exp(\theta \sum_i x_i)$ with different values of $\theta$.  The effect of $\theta\ne 0$ is to give  certain input patterns more weight than others during the training of the BM and QBM.
The BM can correctly classify 104 of the 256 problems (these are the linearly separable problems) for a large range of $\theta$ values (see Figure~\ref{theta_dependence}). 
For $\theta=0$, the QBM can correctly classify 160 of the 256 problems. For $\theta$ in the range 0.1--0.8 the QBM can correctly classify at least 254 problems and for 
$\theta\approx $\ 0.6--0.7, the QBM can correctly classify all 256 problems. Details of the classifications for $\theta=0$ and $\theta=0.7$ are shown in table~\ref{classification}. 

To better understand the strong dependence of the QBM classification accuracy on $\theta$ we list in table~\ref{classification} also the entropy of the eigenvalues of $\rho_\mathrm{qbm}$ and the KL divergence between the ground state 
probability and $q$. We see that  the effect of $\theta>0$ is that the harder problems become closer to a rank one solution and the ground state becomes closer to the target distribution $q$. Why this is the case is not clear and requires more investigation. 
There seems no principled way to choose $\theta$ other than exploring a range of values, but one can clearly think of adaptive schemes such as boosting \cite{freund1997decision}.
The important conclusion is that with a proper $\theta$, the QBM can correctly classify {\em all} 256 problems. 

It is clear that the QBM cannot correctly classify all binary functions for larger problems, because the number of parameters of the QBM scales polynomial as $\cO(n^2)$ while the number of training samples that sets the number of constraints scales exponentially as $2^{n-1}$. Nevertheless, the present study suggests that the QBM with up to second order interaction 
can learn many non linearly separable problems than the BM and it is likely that this also holds for larger problems. 

%\begin{table}
%\bc
%\begin{tabular}{c|c|c|c}
%$n$ & LR/BM & QBM ($\theta=0$) & QBM ($\theta=0.1$)\\\hline
%
%104	& 104	&	104			& 104	\\
%56	& 0 &	56			& 56\\
%96    & 0 & 0				& 94\\\hline
%256 & 104		&	160		& 254
%\end{tabular}
%\ec
%\caption{Summary of classification performance. Table lists the number of problems that can be exactly solved by each method and subdivided into three subsets of problems.
%}
%\label{classification}
%\end{table}

%\begin{figure}
%\bc
%\includegraphics[width=0.4\textwidth]{batch8_all_function4a.eps} %data are batch8_random_function4a.mat
%\includegraphics[width=0.4\textwidth]{batch8_all_function4b.eps} %data are batch8_random_function4b.mat
%\ec
%\caption{Number of classification errors on all $2^8=256$ binary classification problems on $3$ binary input variables for logistic regression (LR), BM and QBM. 
%Problem instances are sorted by increasing LR error. The BM and LR give zero error on the same 104 problems and non-zero error on the remaining 152 problems. 
%The QBM gives zero classification error on 248 problems and non-zero error on the remaining 8 problems.}
%\label{classification}
%\end{figure}

\section{Entanglement in classical data}
\label{entanglement}
The data state Eq.~\ref{eta_classical} $\eta=\ket{\psi}\bra{\psi}$ (with arbitrary phase $\alpha(s)$)  is a much more complex object than the classical probability distribution $q$ from which it is derived. 
It displays full quantum features such as non locality and entanglement.
In this section we explain what these quantum features mean for classical data. 

$\eta$ is a basis independent object that allows measurements in any basis. Measurement in the basis where $\sigma_i^z$ is diagonal yields the classical measurement outcomes. For instance, measuring $\sigma_i^z$ yields outcomes $s_i=\pm 1$ and measuring $\sigma_i^z\sigma_j^z$ yields outcomes $s_is_j=\pm 1$.  The measurement outcome $s_i=\pm 1$ means that the quantum system is in the state $\ket{s_i=\pm 1}$. Repeated measurement yields statistics whose expected values, such as $\av{\sigma_i^z}_\eta=\av{s_i}_q$ or $\av{\sigma_i^z\sigma_j^z}_\eta=\av{s_is_j}_q$ etc., are predicted by the joint probability distribution $q(s_1,\ldots,s_n)$. 

But $\eta$ also allows measurement in any other basis, for instance in the basis where $\sigma_i^x$ is diagonal. The measurement $\sigma^x_i$ yields outcomes $t_i=\pm1 $ meaning that the quantum system is in the state
$\ket{t_i=\pm 1}=\frac{1}{\sqrt{2}}\left(\ket{s_i=1}\pm \ket{s_i=-1}\right)$, which is a superposition of two 'classical' states $\ket{s_i=\pm 1}$. 
Repeated measurement yields statistics whose expected values such as $\av{\sigma_i^x}_\eta$ are given by Eq.~\ref{spin_statistics}. 
Equivalently, one can define a probability distribution $\tilde{q}$ in the new basis $\ket{t}=\ket{t_1,\ldots,t_n}$ and $\av{\sigma_i^x}_\eta=\sum_{t_i} t_i \tilde{q}(t_i)$. 
In general, measurements that are in an orthogonal basis $\ket{t}=\sum_s U_{ts}\ket{s}$, with $U$ a unitary transformation are described by the classical distribution $\tilde{q}(t)=|\av{t|\psi}|^2$
(see~\ref{projective}).
 
The difference between $q$ and $\eta$ is reflected in the difference between the classical mutual information $I_c(q)$ and the 
quantum mutual information $I(\eta)$ between two sub systems $A$ and $B$ (see~\ref{MI}). 
For a pure state, such as the data state $\eta$,
the total entropy is zero $h(\eta)=0$ and the entropies of the sub systems are equal: $h(\eta_A)=h(\eta_B)$ (see~\ref{schmidt}). 
\footnote{Note, that this identity holds independent of the sizes of sub systems $A$ and $B$: $A$ can be a single bit and $B$ the rest of the universe. }
Thus, the quantum mutual information Eq.~\ref{I} is
\bea
I(\eta)=h(\eta_A)+h(\eta_B)-h(\eta)=2h(\eta_A)\label{Iq}
\eea
For pure states, the quantum mutual information is equal to the entanglement \cite{maziero2009classical}. 

{\em The quantum mutual information $I(\eta)$ in the data state $\eta$ is never less than the classical mutual information $I_c(q)$ of the distribution $q$ from which it is composed, ie. $I_c(q)\le I(\eta)$.}
The proof is as follows. 
The diagonal of $\eta$ is the classical distribution $q$. The operation that maps a density matrix to its diagonal, is known as the incoherence operator $\Pi(\eta)=\eta^\mathrm{diag}=q$. The incoherence operator is a completely positive trace preserving (CPTP) operation. For any CPTP operation $\Phi$  \cite{lindblad1975completely}
\bea
S(\Phi(\rho),\Phi(\sigma))\le S(\rho,\sigma) \label{monotonicity_cptp}
\eea
Thus we establish the bound: 
\beaa
I_c(q)=S(q,q_A q_B)= S(\Pi(\eta),\Pi(\eta_A)\Pi(\eta_B))\le S(\eta,\eta_A\eta_B)=I(\eta) 
\eeaa
which completes the proof.

\begin{figure}
\bc
\subfigure[
$\alpha(s)$ is random.]
{\includegraphics[width = 0.4\textwidth]{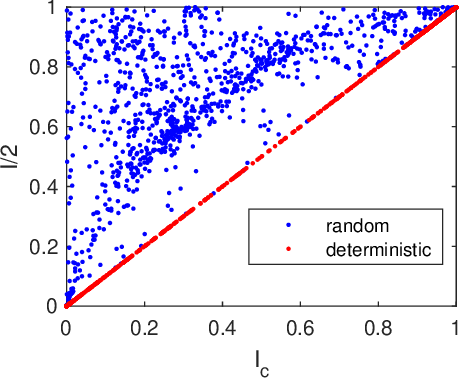}\label{file19a_left}}
\subfigure[
$\alpha(s)=0$.]
{\includegraphics[width = 0.4\textwidth]{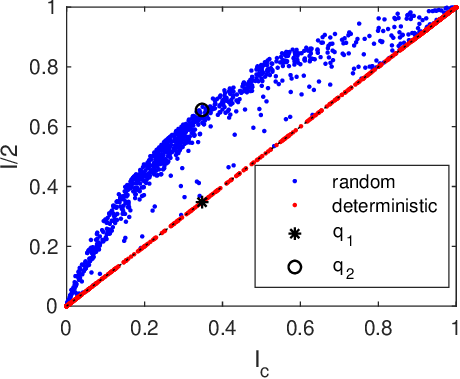}\label{file19a_right}}
\ec
\caption{Scatter plot of $C=I(\eta)/2$  versus $I_c(q)$ between sub systems $A$ and $B$ for three binary variables $A=(s_1,s_2)$ and $B=s_3$ with random probability distribution $q(s_1,s_2,s_3)$.  Blue: 1000 random distributions are generated, with each component $q(s_1,s_2)$ generated independently from a Beta distribution with parameters $\alpha=\beta=0.1$. Subsequently, $q$ is normalized.  Red: The same, but with a deterministic relation $s_3=s_2$. (a) $\eta$ is constructed with random phases $\alpha(s)$; (b) $\eta$ is constructed with phases $\alpha(s)=0$.}
\label{file19a}
\end{figure}

{\em In fact, we conjecture the stronger statement that $I_c(q)\le I(\eta)/2$ but we have not been able to prove this. }
We illustrate this with a numerical example. Consider three spins
$s_1,s_2,s_3$ and define sub system $A=(s_1,s_2)$ and $B=s_3$. We
generate random probability distributions $q$ on these three spins and
compute $I_c(q)$ and $I(\eta)$ using Eq.~\ref{eta_classical} with
random phases $\alpha(s)$ (Figure~\ref{file19a_left} blue dots) as well as with $\alpha(s)=0$ 
(fig~\ref{file19a_right} blue dots). We observe that in all cases $ I_c(q) \le I(\eta)/2$. 
When the states of subsystem $B$ are a deterministic function of the states of $A$ we can show $I_c(q)=I(\eta)/2$ (see~\ref{cc} for the proof and illustrated in Figure~\ref{file19a} red dots). 

One can write for a pure state  that  $I(\eta)=C+Q$ with $C=Q=I(\eta)/2$ \cite{groisman2005quantum,maziero2009classical,vedral2017foundations}.
$C$ is called the (maximal) classical correlation (see~\ref{cc}) and $Q$ is called the quantum correlation.  
Measuring in an orthogonal basis in which the reduced density matrices of the two sub systems are diagonal defines a distribution $\tilde{q}$ between measurement outcomes 
for which $I_c(\tilde{q})=C\ge I_c(q)$. See~\ref{cc} for details.
We illustrate this for $\alpha(s)=0$ with the distributions $q_1$ and $q_2$, defined as  
\beaa
q_1(s_2,s_3)=\left(\begin{tabular}{cc}$0.0652 $& $0$\\$0$& $0.9348$\end{tabular}\right)\qquad 
q_2(s_2,s_3)=\left(\begin{tabular}{cc}$0.25 $& $0.375$\\$0.375$& $0$\end{tabular}\right)
\eeaa
independent of $s_1$, that are chosen such that the classical mutual information between $A$ and $B$ is identical: $I_c(q_1)=I_c(q_2)$. In $q_1$, $s_B$ depends deterministically on $s_A$ and $I_c(q_1)=I(\eta_1)/2$ ($*$ in Figure~\ref{file19a_right}). In $q_2$, the relation is not deterministic and $I_c(q_2)< I(\eta_2)/2$ ($o$ in Figure~\ref{file19a}). Measurement in the orthogonal basis defined by the Schmidt decomposition yields 
\beaa
\tilde{q_2}(t_2,t_3)=\left(\begin{tabular}{cc}$ 0.8307$& $0$\\$0$& $0.1693$\end{tabular}\right) \qquad U_A(t_2,s_2)=\left(\begin{tabular}{cc}$ 0.83$& $0.5577$\\$0.5577$& $-0.83$\end{tabular}\right) 
\eeaa
and $U_B=U_A$ and $I_c(\tilde{q}_2)=I(\eta_2)/2> I_c(q_2)$. 

%The point is illustrated for the parity problem on three variables with $A=s_1$ and $B=s(s_2,s_3)$. We write $q(s_A,s_B)$ in matrix form
%\beaa
%q=\frac{1}{4}\left(\begin{tabular}{cccc}$1$& $0$ &$0$&$1$\\$0$& $1$ &$1$&$0$\end{tabular}\right)
%\eeaa
%The SVD of $\sqrt{q}=USV^T$ yields %file19.d
%\beaa
%U=\frac{1}{\sqrt{2}}\left(\begin{tabular}{cc}$1$& $1$\\$1$& $-1$\end{tabular}\right)\quad
%S=\frac{1}{\sqrt{2}}\left(\begin{tabular}{cccc}$1$& $0$ &$0$&$0$\\$0$& $1$&$0$&$0$\end{tabular}\right)\quad
%V=\left(\begin{tabular}{cccc}
%$\frac{1}{2}$& $\frac{1}{2}$ &$0$&$-\frac{1}{\sqrt{2}}$\\
%$\frac{1}{2}$& $-\frac{1}{2}$ &$-\frac{1}{\sqrt{2}}$&$0$\\
%$\frac{1}{2}$& $-\frac{1}{2}$ &$\frac{1}{\sqrt{2}}$&$0$\\
%$\frac{1}{2}$& $\frac{1}{2}$ &$0$&$-\frac{1}{\sqrt{2}}$\
%\end{tabular}\right)
%\eeaa
%and $\tilde{q}(t_A,t_B)=S(t_A,t_B)^2$. 

Fig.~\ref{file19a} suggests that the quantum implementation of the classical distribution Eq.~\ref{eta_classical} using non-zero phases $\alpha(s)$ can be superior than using $\alpha(s)=0$ because for the same classical mutual information $I_c(q)$ it yields larger quantum mutual information $I(\eta)$. Therefore, an optimized orthogonal measurement can yield more classical mutual information $I_c(\tilde{q})=I(\eta)/2$ for properly chosen non-zero $\alpha(s)$ than for $\alpha(s)=0$.

\subsection{Non orthogonal measurement and the Bell inequality}\label{bell}

The statistics of local orthogonal quantum measurements are described by a
classical joint probability model $\tilde{q}$ and the mutual
information is restricted to $I_c(\tilde{q})\le C=I(\eta)/2$. 
The implication is that the remaining quantum information $Q$ resides in non-local features of the state $\eta$ \cite{maziero2009classical}. This relates to the fact that the statistics of non-orthogonal measurements violate the Bell inequality, meaning that they cannot be described by a classical probability model involving 
local variables. Instead such measurement statistics require a description in terms of non-local variables (see~\ref{non-local}).

We illustrate the excess quantum information $Q$ and the violation of the Bell inequality for the simplest possible example of two fully correlated binary variables  $s=(s_1,s_2)$ with joint probability distribution 
\bea
q(s_1,s_2)=\frac{1}{2}\left(\begin{tabular}{cc}1 & 0\\0& 1\end{tabular}\right)\qquad \psi(s_1,s_2)=
\frac{1}{\sqrt{2}}\left(\begin{tabular}{cc}$e^{i\alpha_1}$ & 0\\0& $e^{i\alpha_2}$\end{tabular}\right)\label{belldata}
\eea
where $\psi(s)$ is given by Eq.~\ref{eta_classical}. 

Note, that the classical mutual information between the two bits is $I_c(q)=1$ and the quantum mutual information is $I(\eta)=2$ (Eq.~\ref{Iq}). Half of the quantum mutual information $C=I(\eta)/2=I_c(q)$ is the classical correlation. The other half $Q$ is the quantum correlation. \cite{groisman2005quantum} provides an intuitive explanation of this mysterious extra bit of information using the argument of bit erasure. Assume $\alpha_1=\alpha_2=0$. Suppose that Alice has access to spin 1 and Bob has access to spin 2. Alice can apply to her spin one of two unitary
transformations $1$ or $\sigma^z$ with equal probability. It is easy to show that this operation transforms $\eta=\ket{\psi}\bra{\psi}$ to a diagonal density matrix $\eta'=\text{diag}(q)$. The mutual information $I(\eta')=1$ since the bits are fully correlated in the classical sense. 
Thus, by erasing one bit, which is forgetting the information which unitary Alice has applied, one bit of mutual information is lost. Alice can apply to her spin again one of two unitary
transformations $1$ or $\sigma^x$ with equal probability. This operation either flips or not flips her spin. The final state $\eta''=\eta_1\otimes \eta_2$ is independent and the mutual information $I(\eta'')=0$. Thus, the explanation why $I(\eta)=2$ is that it takes the erasure of two bits of information to remove the mutual information between $A$ and $B$.

We now show how $\eta$ violates the Bell inequality. 
We consider 4 measurements $M_{ij}=A_i\otimes B_j,i,j=1,2$ 
\beaa
A_1&=&\sigma_1^z,\qquad A_2=\sigma_1^x \sin \theta +\sigma_1^z \cos \theta \\
 B_1&=&\sigma_2^z\qquad B_2=\sigma_2^x \sin \phi +\sigma_2^z \cos \phi \label{bell_operators}
\eeaa
Denote the measurement outcomes as $m_{ij}=(a_{ij},b_{ij}) ,i,j=1,2$ with $a_{ij}, b_{ij} =\pm 1$. The locality assumption states that for instance, the outcome of measurement $A_1$ does not depend on whether 
it is combined with the measurement $B_1$ or $B_2$. In other words, $m_{ij}=(a_i,b_j)$. This implies that the measurement outcomes can be described by a probability distribution $p(a_1,b_1,a_2,b_2)$ on 4 binary variables. 

\begin{figure}
\bc
\includegraphics[width=0.4\textwidth]{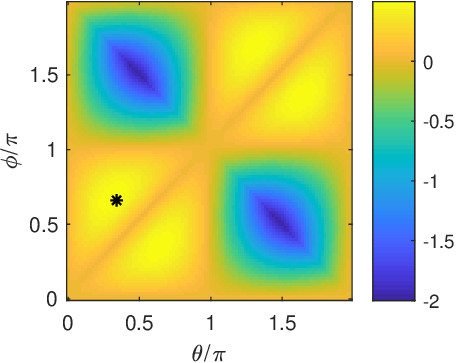}
\ec
\caption{Violation of the Bell inequality Eq.~\ref{bell1} in classical data. The quantum state is built from classical data of two fully correlated binary variables Eq.~\ref{belldata}. The figure shows $B(\theta,\phi)$ (Eq.~\ref{bell1}) as a function of $\theta,\phi$. 
These measurements maximally violate the locality assumption for $(\theta,\phi)=\left(\frac{\pi}{3},\frac{2\pi}{3}\right)$ indicated by *.}
%generated with file13.m
\label{bell3}
\end{figure}

Note, that in the state $\ket{\psi}$ the measurement outcomes $a_1=b_1$ are fully correlated. Thus,  we can discard $b_1$ since it adds no additional information. 
%Denote $a=a_1, b=a_2, c=b_2$ and 
%\bea
%p(a_1,b_1,a_2,b_2)=p(a,b,c)\label{pbell}
%\eea
Since the measurement outcomes $a_1,a_2,b_2=\pm 1$ they trivially satisfy $a_1 a_2 +a_1 b_2-a_2b_2\le 1$ as well as any of the permutations of $a_1,a_2,b_2$ \cite{cerf1997entropic}. Let us assume that there exist a classical probability distribution $p(a_1,a_2,b_2)$ that describes the outcome statistics. Taking the expectation with respect to $p$ we obtain the 
Bell inequality 
\bea
B=|\av{a_1a_2}-\av{a_1b_2}|+\av{a_2b_2}-1\le 0 \label{bell1}
\eea
We compute the quantum statistics in the state $\psi$:
\footnote{Note, that $A_1$ and $A_2$ do not commute and $A_1A_2$ is not Hermitian. Therefore the quantum expectation $\av{A_1A_2}$ is not  real while the classical correlation $\av{a_1a_2}$ are real. We therefore identify the classical correlations with the correlations of the Hermitian operator $\frac{1}{2}(A_1A_2+A_2A_1)$.}
\beaa
\av{a_1a_2}&=&\sum_{s_1's_2',s_1,s_2}  \psi^*(s_1',s_2')\frac{1}{2}\left(A_1A_2+A_2A_1\right)_{s_1',s_1} \psi(s_1,s_2)=\cos\theta\\
\av{a_1b_2}&=&\sum_{s_1's_2',s_1,s_2} \psi^*(s_1',s_2') \left(A_1\right)_{s_1',s_1} \left(B_2\right)_{s_2',s_2} \psi(s_1,s_2)=\cos\phi\\
\av{a_2b_2}&=&\sum_{s_1's_2',s_1,s_2} \psi^*(s_1',s_2') \left(A_2\right)_{s_1',s_1} \left(B_2\right)_{s_2',s_2} \psi(s_1,s_2)\\
&=&\cos(\theta)\cos(\phi)+\sin(\theta)\sin(\phi)\cos(\alpha_1-\alpha_2)
\eeaa
Substituting these correlations in Eq.~\ref{bell1}, we observe that $B$ is maximized when $\alpha_1-\alpha_2=0,\pi$ and $\max_{\theta,\phi,\alpha_1-\alpha_2} B=\frac{1}{2}$ which violates the Bell inequality.  $\theta,\phi$ can take multiple values as is illustrated for $\alpha_1-\alpha_2=0$ in  Figure~\ref{bell3} where we plot $B$ as a function of $\theta,\phi$. Note, that these measurements are non-orthogonal. This example demonstrates that the statistics of non-orthogonal measurements cannot be described by a probability distribution of local outcomes. 

This seems to contradict the fact that $\eta$ is built from a classical distribution $q$ that has only local variables. However, there is no contradiction because the Bell condition assumes that the quantum correlations are linear in the classical distribuion $p(a_1,a_2,b_2)$, while in fact they are quadratic in $\sqrt{q}$ (Eqs.~\ref{avA} and~\ref{spin_statistics}). 
%The Bell construction can be generalized to any state $\psi=\sum_k a_k v_k w_k^\dagger$ with $v_k$ and $w_k$ orthogonal states of the two sub systems, when at least two of the $a_k$ are nonzero \cite{wigner1970hidden}. 
%
%

\section{Discussion}

In this paper, we have addressed the problem how to represent a classical data distribution in a quantum system. 
The proposed method is to learn quantum Hamiltonian that is such that its ground state approximates the given classical distribution. 
The essential new ingredient is to represent a classical distribution as a rank one density matrix such that $\text{diag}(\eta)=q$. In particular, we proposed Eq.~\ref{eta_classical}. 

This choice is not unique. First of all, one needs to specify the unitary transformation $U(\alpha)=\text{diag}(e^{i\alpha})$ in Eq.~\ref{eta_classical}.  In our numerical experiments in section~\ref{classical} we have assumed $\alpha=0$. The choice of $\alpha$ affects the learning and raises the question how to best choose the phases $\alpha$. Part of the answer is to observe that $U(\alpha)$ can be equivalently considered part of a model $U^\dagger \rho U$. Since for an arbitrary data set there is no principled choice of $\rho$, one can extend this reasoning to (a parametrized model) for $U$. $U$ and $\rho$ can then be jointly optimized to maximize $L$. 
The results on entanglement in Figure~\ref{file19a} suggest that choosing $\alpha(s)$ non-zero may result in a quantum state that has more entanglement than when $\alpha(s)=0$. The issue of the choice of $\alpha(s)$ has not been further explored and is left for future research. 

Secondly, a further generalization is to define 
$\eta =\mu \eta^*+ (1-\mu) \text{diag}(q)$ with $0\le \mu \le 1$ and $\eta^*$ the rank one density matrix Eq.~\ref{eta_classical}. Note that when $q(s)>0$ for all $s$ and $\mu\ne 1$, $\eta$ is a positive matrix while $\eta^*$ is not (it has eigenvalues 0). 
Thus choosing $\eta$ as a learning target with $\mu \ne1$ in the quantum likelihood Eq.~\ref{Lquantum} instead of $\eta^*$ may improve convergence speed since $S(\eta,\rho)$ is strictly convex while $S(\eta^*,\rho)$ is not \cite{carlen2010trace}. 
On the other hand, an advantage of the choice $\eta^*$, instead of $\eta$  is that it maximizes the entanglement and quantum mutual information: $I(\eta^*)\ge I(\eta)$ for all $\mu$ and any sub division $A,B$. 
This follows from Eq.~\ref{monotonicity_cptp} and the fact that the mapping $\Phi: \eta^* \to \eta$  is a CPTP operation and $\Phi(\eta_A\otimes \eta_B)=\Phi(\eta_A)\otimes\Phi(\eta_B)$. Thus, the gain in quantum mutual information is largest for $\mu=1$. 
The set of density matrices $\eta$ that have diagonal $q$ is much larger than what we
have considered in this paper and other choices may be considered.

When transforming a probability distribution to a density matrix by
Eq.~\ref{eta_classical}, the mutual information between subsets of
variables increases because $I(\eta)\ge 2 I_c(q)$. $C=I(\eta)/2$ is the so-called (maximal) classical correlations and the extra mutual information
$I(\eta)/2-I_c(q)$ is accessible by optimizing the orthogonal measurement basis. Its outcome statistics are described by a classical probability distribution on local variables. 
When implementing this idea in a quantum device, these extra correlations can in principle be used to improve performance in machine learning or other applications. 

The remaining quantum mutual information $Q=I(\eta)/2$ is 'non local'  and cannot be accessed by local orthogonal measurements.  This part of the mutual information is the true quantum entanglement. The non-locality is manifested in the violation of the Bell inequality and occurs even for classical distributions $q$ that contain deterministic relations, such as two fully correlated classical spins or the parity problem that we considered in Figure~\ref{parity_bm}. It is an open and interesting question whether these non-local features can be used in a quantum device for the type of learning problems that we considered here.

For large simulations, learning the QBM is intractable. In each learning iteration one must compute statistics $\av{H_r}_\rho$ for the current estimated density matrix $\rho$. In principle, it requires $\cO\left(2^{2n}\right)$ operations and memory to compute the entire density matrix. To generate the results of Figure~\ref{file4_afh} we effectively made use of a low rank approximation using $L=6$ extreme eigenvectors and a sparse representation, requiring $\cO\left(L 2^{n}\right)$ computation.  Clearly, this is still exponential in the number variables and does not scale to large problem instances.

As in classical BM learning, one can apply various approximate inference methods to estimate $\av{H_r}_\rho$. One promising approach is to
approximate $\av{H_r}_\rho$ by its ground state statistics. This can be done by computing in each learning iteration an approximation to the ground state wave function by 
making a variational Anzats and minimize the Raleigh quotient for the current instance of the Hamiltonian \cite{mcmillan1965ground, carlson2015quantum}. In particular, 
\cite{carleo2017solving} use a (classical) restricted Boltzmann machine as the variational Anzats for the wave function that shows great potential. This approach is currently under investigation.
%In general, when the density matrix has finite temperature and is not a rank one matrix, one may need to estimate the statistics using diffusion Monte Carlo \cite{ceperley1986quantum,ceperley1995path} or mean field methods \cite{matera2010evaluation,dominguez2018quantum}.

Alternatively, one can imagine that the computation of the quantum statistics can be done on physical quantum hardware. This would give a hybrid quantum-classical computing scheme
\cite{preskill2018quantum}, where the learning iterations are executed on a classical computer and the computation of $\av{H_r}_\rho$ for the current Hamiltonian (or its ground state approximation) would be done on the quantum device.  Initial encouraging results were obtained to implement the approximate quantum learning rule \cite{amin2016quantum} using quantum annealing on a D-Wave computer \cite{korenkevych2016benchmarking}. See also \cite{mitarai2018quantum}. An obvious next step is to use the quantum learning rules proposed here to improve these results. 

%It may be possible that using more general positive operator valued measurements (POVM) one can exceed $I/2$. This question is left for future research.

%The data state $\eta$ corresponds to a positive wave function (in the $\sigma^z$ basis). This relates the QBM learning problem to the learning of stoquastic Hamiltonians, since their 
%ground states are positive. Computing quantum statistics of stoquastic Hamiltonians is much easier than for general Hamiltonians because they have no sign problem \cite{bravyi2006complexity}. 
%It is therefore interesting to restrict the learning problem to stoquastic Hamiltonians. 

%The definition of the data wave function $\psi(s)=\sqrt{q(s)}$ can be generalized to $\psi(s)=e^{i\phi(s)}\sqrt{q(s)}$. Any choice of $\phi(s)$ is in principle equally valid and may affect the subsequent learning. We saw an example of this in Figure~\ref{file4_afh} where in the case of the AFH model, the QBM ground state wave function only differs from the classical solution $\sqrt{p_\mathrm{bm}}$ in the sign of some components, but their quantum statistics are very different. 
%However, the data do not provide any information on the choice of $\phi(s)$. Therefore, one may consider to optimize $\phi(s)$ as part of the learning process. 
%

%The entanglement between $A$ and $B$ can be defined in different ways \cite{vedral1998entanglement,vrehavcek2003quantification}. 

\subsection{Acknowledgments}
We like to thank the anonymous reviewers for useful comments on the earlier version of this manuscript. This research was funded in part by ONR Grant N00014-17-1-2569. 

\vspace{5mm}

\bibliography{/Users/bertkappen/doc/authors}

\begin{thebibliography}{10}

\bibitem{kieferova2017tomography}
M{\'a}ria Kieferov{\'a} and Nathan Wiebe.
\newblock Tomography and generative training with quantum boltzmann machines.
\newblock {\em Physical Review A}, 96(6):062327, 2017.

\bibitem{amin2016quantum}
Mohammad~H Amin, Evgeny Andriyash, Jason Rolfe, Bohdan Kulchytskyy, and Roger
  Melko.
\newblock Quantum boltzmann machine.
\newblock {\em Physical Review X}, 8(2):021050, 2018.

\bibitem{lecun2015deep}
Yann LeCun, Yoshua Bengio, and Geoffrey Hinton.
\newblock Deep learning.
\newblock {\em Nature}, 521(7553):436--444, 2015.

\bibitem{mnih2015human}
Volodymyr Mnih, Koray Kavukcuoglu, David Silver, Andrei~A Rusu, Joel Veness,
  Marc~G Bellemare, Alex Graves, Martin Riedmiller, Andreas~K Fidjeland, Georg
  Ostrovski, et~al.
\newblock Human-level control through deep reinforcement learning.
\newblock {\em Nature}, 518(7540):529--533, 2015.

\bibitem{silver2016mastering}
David Silver, Aja Huang, Chris~J Maddison, Arthur Guez, Laurent Sifre, George
  van~den Driessche, Julian Schrittwieser, Ioannis Antonoglou, Veda
  Panneershelvam, Marc Lanctot, et~al.
\newblock Mastering the game of go with deep neural networks and tree search.
\newblock {\em Nature}, 529(7587):484--489, 2016.

\bibitem{kadowaki1998quantum}
Tadashi Kadowaki and Hidetoshi Nishimori.
\newblock Quantum annealing in the transverse ising model.
\newblock {\em Physical Review E}, 58(5):5355, 1998.

\bibitem{heim2015quantum}
Bettina Heim, Troels~F R{\o}nnow, Sergei~V Isakov, and Matthias Troyer.
\newblock Quantum versus classical annealing of ising spin glasses.
\newblock {\em Science}, 348(6231):215--217, 2015.

\bibitem{adachi2015application}
Steven~H Adachi and Maxwell~P Henderson.
\newblock Application of quantum annealing to training of deep neural networks.
\newblock {\em arXiv preprint arXiv:1510.06356}, 2015.

\bibitem{benedetti2016estimation}
Marcello Benedetti, John Realpe-G{\'o}mez, Rupak Biswas, and Alejandro
  Perdomo-Ortiz.
\newblock Estimation of effective temperatures in quantum annealers for
  sampling applications: A case study with possible applications in deep
  learning.
\newblock {\em Physical Review A}, 94(2):022308, 2016.

\bibitem{carleo2017solving}
Giuseppe Carleo and Matthias Troyer.
\newblock Solving the quantum many-body problem with artificial neural
  networks.
\newblock {\em Science}, 355(6325):602--606, 2017.

\bibitem{carrasquilla2017machine}
Juan Carrasquilla and Roger~G Melko.
\newblock Machine learning phases of matter.
\newblock {\em Nature Physics}, 13(5):431, 2017.

\bibitem{biamonte2017quantum}
Jacob Biamonte, Peter Wittek, Nicola Pancotti, Patrick Rebentrost, Nathan
  Wiebe, and Seth Lloyd.
\newblock Quantum machine learning.
\newblock {\em Nature}, 549(7671):195, 2017.

\bibitem{cerf1997negative}
Nicolas~J Cerf and Chris Adami.
\newblock Negative entropy and information in quantum mechanics.
\newblock {\em Physical Review Letters}, 79(26):5194, 1997.

\bibitem{cerf1999quantum}
Nicolas~J Cerf and Christoph Adami.
\newblock Quantum extension of conditional probability.
\newblock {\em Physical Review A}, 60(2):893, 1999.

\bibitem{schack2001quantum}
Ruediger Schack, Todd~A Brun, and Carlton~M Caves.
\newblock Quantum bayes rule.
\newblock {\em Physical Review A}, 64(1):014305, 2001.

\bibitem{wiebe2019generative}
Nathan Wiebe and Leonard Wossnig.
\newblock Generative training of quantum boltzmann machines with hidden units.
\newblock {\em arXiv preprint arXiv:1905.09902}, 2019.

\bibitem{wiersema2019implementing}
RC~Wiersema and HJ~Kappen.
\newblock Implementing perceptron models with qubits.
\newblock {\em Physical Review A}, 2019.
\newblock In press, arXiv:1905.06728.

\bibitem{ack85}
D.~Ackley, G.~Hinton, and T.~Sejnowski.
\newblock A learning algorithm for {B}oltzmann {M}achines.
\newblock {\em Cognitive Science}, 9:147--169, 1985.

\bibitem{carlen2010trace}
Eric Carlen.
\newblock {\em Trace inequalities and quantum entropy: an introductory course},
  volume 522, pages 73--140.
\newblock American Mathematical Society, 2010.

\bibitem{torlai2017many}
Giacomo Torlai, Guglielmo Mazzola, Juan Carrasquilla, Matthias Troyer, Roger
  Melko, and Giuseppe Carleo.
\newblock Many-body quantum state tomography with neural networks.
\newblock {\em arXiv preprint arXiv:1703.05334}, 2017.

\bibitem{ferrie2014self}
Christopher Ferrie.
\newblock Self-guided quantum tomography.
\newblock {\em Physical review letters}, 113(19):190404, 2014.

\bibitem{granade2017practical}
Christopher Granade, Christopher Ferrie, and Steven~T Flammia.
\newblock Practical adaptive quantum tomography.
\newblock {\em New Journal of Physics}, 19(11):113017, 2017.

\bibitem{tkavcik2014searching}
Ga{\v{s}}per Tka{\v{c}}ik, Olivier Marre, Dario Amodei, Elad Schneidman,
  William Bialek, and Michael~J Berry~II.
\newblock Searching for collective behavior in a large network of sensory
  neurons.
\newblock {\em PLoS computational biology}, 10(1):e1003408, 2014.

\bibitem{freund1997decision}
Yoav Freund and Robert~E Schapire.
\newblock A decision-theoretic generalization of on-line learning and an
  application to boosting.
\newblock {\em Journal of computer and system sciences}, 55(1):119--139, 1997.

\bibitem{maziero2009classical}
J~Maziero, L~C\_ Celeri, RM~Serra, and V~Vedral.
\newblock Classical and quantum correlations under decoherence.
\newblock {\em Physical Review A}, 80(4):044102, 2009.

\bibitem{lindblad1975completely}
G{\"o}ran Lindblad.
\newblock Completely positive maps and entropy inequalities.
\newblock {\em Communications in Mathematical Physics}, 40(2):147--151, 1975.

\bibitem{groisman2005quantum}
Berry Groisman, Sandu Popescu, and Andreas Winter.
\newblock Quantum, classical, and total amount of correlations in a quantum
  state.
\newblock {\em Physical Review A}, 72(3):032317, 2005.

\bibitem{vedral2017foundations}
Vlatko Vedral.
\newblock Foundations of quantum discord.
\newblock In {\em Lectures on General Quantum Correlations and their
  Applications}, pages 3--7. Springer, 2017.

\bibitem{cerf1997entropic}
N.~J. Cerf and C.~Adami.
\newblock Entropic bell inequalities.
\newblock {\em Physical Review A}, 55(5):3371, 1997.

\bibitem{mcmillan1965ground}
William~Lauchlin McMillan.
\newblock Ground state of liquid he 4.
\newblock {\em Physical Review}, 138(2A):A442, 1965.

\bibitem{carlson2015quantum}
J~Carlson, Stefano Gandolfi, Francesco Pederiva, Steven~C Pieper, Rocco
  Schiavilla, KE~Schmidt, and Robert~B Wiringa.
\newblock Quantum monte carlo methods for nuclear physics.
\newblock {\em Reviews of Modern Physics}, 87(3):1067, 2015.

\bibitem{preskill2018quantum}
John Preskill.
\newblock Quantum computing in the nisq era and beyond.
\newblock {\em Quantum}, 2:79, 2018.

\bibitem{korenkevych2016benchmarking}
Dmytro Korenkevych, Yanbo Xue, Zhengbing Bian, Fabian Chudak, William~G
  Macready, Jason Rolfe, and Evgeny Andriyash.
\newblock Benchmarking quantum hardware for training of fully visible boltzmann
  machines.
\newblock {\em arXiv preprint arXiv:1611.04528}, 2016.

\bibitem{mitarai2018quantum}
Kosuke Mitarai, Makoto Negoro, Masahiro Kitagawa, and Keisuke Fujii.
\newblock Quantum circuit learning.
\newblock {\em Physical Review A}, 98(3):032309, 2018.

\bibitem{araki2002entropy}
Huzihiro Araki and Elliott~H Lieb.
\newblock Entropy inequalities.
\newblock In {\em Inequalities}, pages 47--57. Springer, 2002.

\bibitem{miszczak2011singular}
Jaros{\l}aw~Adam Miszczak.
\newblock Singular value decomposition and matrix reorderings in quantum
  information theory.
\newblock {\em International Journal of Modern Physics C}, 22(09):897--918,
  2011.

\bibitem{modi2012classical}
Kavan Modi, Aharon Brodutch, Hugo Cable, Tomasz Paterek, and Vlatko Vedral.
\newblock The classical-quantum boundary for correlations: discord and related
  measures.
\newblock {\em Reviews of Modern Physics}, 84(4):1655, 2012.

\bibitem{wigner1970hidden}
Eugene~P Wigner.
\newblock On hidden variables and quantum mechanical probabilities.
\newblock {\em American Journal of Physics}, 38(8):1005--1009, 1970.

\bibitem{bell1964einstein}
John~S Bell.
\newblock On the einstein podolsky rosen paradox.
\newblock {\em Physics Physique Fizika}, 1(3):195, 1964.

\bibitem{bell2001problem}
John~S Bell.
\newblock On the problem of hidden variables in quantum mechanics.
\newblock In {\em John S Bell On The Foundations Of Quantum Mechanics}, pages
  1--6. World Scientific, 2001.

\end{thebibliography}
\bibliographystyle{unsrt}
\appendix
\section{QBM learning using the classical likelihood}
\label{appendix:amin}
\cite{amin2016quantum} and one of the methods discussed in \cite{kieferova2017tomography} use the classical likelihood for learning the QBM by considering the diagonal of the density matrix.
Write $\rho(s,s')=\delta_{s,s'}p(s)$. 
For each state $s$, define $\Lambda_s$ a matrix with components $\Lambda_s(s',s'')=\delta_{s,s'}\delta_{s,s''}$.
Then $p(s)=\Tr\left( \Lambda_s \rho\right)$ and the classical log likelihood is
\bea
L=\sum_s q(s)\log \Tr\left( \Lambda_s \rho\right)=\sum_s q(s)\log \Tr\left( \Lambda_s e^H\right)-\log Z\label{amin}
\eea
Note, that this expression differs from the quantum likelihood MT Eq.~4 in the first term only. Its
gradient is given by
\beaa
\frac{\partial}{\partial w_r}\log \Tr\left(\Lambda_s e^H\right)=\frac{1}{\Tr\left(\Lambda_s e^H\right)}\int_0^1 dt \Tr\left( \Lambda_s e^{Ht}H_r e^{H(1-t)}\right)
\eeaa
Because of $\Lambda_s$ does not commute with $e^{Ht}$, the time integration remains and the expression cannot be easily evaluated. \cite{amin2016quantum} address this problem by deriving a lower bound on their likelihood using the Golden-Thompson inequality and maximizing this bound. However, as the authors admit,  this procedure is clearly sub optimal and is inconsistent when learning transverse field components.

\section{Quantum Boltzmann machine details\label{qbm}}
In the Hamiltonian Eq.~\ref{Hamiltonian},
$\sigma_i^{x,y,z}$ are $2\times 2$ Pauli spin matrices in the $\sigma^z$ basis. Then $\sigma_i^z=\mathrm{diag}(1,-1)$, its eigenvectors are the two component unit vectors $(1,0)$ and $(0,1)$, which we denote as $\ket{s_i=\pm1}$ and eigenvalues $s_i=\pm 1$, respectively. On this basis
\beaa
\sigma_i^z \ket{s_i}=s_i \ket{s_i}, \qquad \sigma_i^x \ket{s_i}=\ket{-s_i}\qquad \sigma_i^y\ket{s_i}=is_i \ket{-s_i}
\eeaa
For $n$ spins, the basis is the tensor product of the basis vectors $\ket{s_i=\pm 1}$ denoted as $\ket{s}=\ket{s_1,\ldots,s_n}$. On this basis, 
$H$ is a $2^n\times 2^n$ matrix with matrix elements 
\beaa
\av{s'|H|s}&=&\sum_{i=1}^n (w_i^x +iw_i^ys_i)\delta_{s',F_i s} +\sum_{i=1,j>i}^n \left(w_{ij}^{x} -w_{ij}^ys_i^k s_j^k\right)\delta_{s',F_iF_js} \\
&+&\left(\sum_{i=1}^n w_i^z s_i+\sum_{i<j}^n w_{ij}^z s_is_j\right)\delta_{s',s}
\eeaa
with $F_is$ the state $s$ with spin $i$ flipped to $-s_i$ and all other spins unchanged.

\section{Details classification performance}
Details of classification performance for all 256 classification problems on 3 binary inputs are summarized in table~\ref{classification}.
\ben
\item A subset of 104 problems is linearly separable (each problem has classification error 0) and can be classified exactly by all methods. The remaining problems are not linearly separable.
\item A subset of 56 problems cannot be classified by BM (each problem has classification error 2-4) and can be exactly classified by QBM. 
\item A subset of 88 problems. For $\theta=0$ the QBM solution is approximately rank two, with eigenvalues $\lambda\approx (0.75, 0.25)$ and cannot correctly classify these problems. For $\theta=0.7$ the QBM solution is closer to rank one and these problems can be exactly classified by QBM. 
\item A subset of 8 problems that contain the parity and parity-like problems. For these problems, the solution $\rho_\mathrm{qbm}$ is rank two with two eigenvalues $\lambda=0.5$. 
For $\theta=0$, the ground state of the Hamiltonian is degenerate and there is no solution. For $\theta=0.7$ the symmetry is broken and these problems can be exactly classified by QBM. 
\een

As a sanity check, we also computed the logistic regression solution $p_\text{lr}(y|x)=\sigma\left(y\sum_i w_i x_i\right)$ that maximizes the conditional classical likelihood $\sum_{x,y} q(x)q(y|x) \log \sigma\left(y \sum_i w_i x_i\right)$ instead of the joint classical likelihood that is used for the BM.  We found no significant quality differences between the logistic regression and BM solutions (data not shown).
We observe that with $\theta=0.7$, the QBM solution is always closer to a rank one solution than for $\theta=0$. For $\theta=0.7$, the quality of the ground state approximation as measured by the KL divergence is excellent for the subsets of 8 problems for which the $\theta=0$ solution is degenerate, while not much worse for the other problems. 

% th_org=0
%104 lr 0 0 bm 0 0 qbm 0 0 qbm1 0 0 ent 0.00 0.01 Sqbm 8.65e-09 3.00e-03 kl_q -8.60e-16 8.21e-06
%56 lr 2 4 bm 2 4 qbm 0 0 qbm1 0 0 ent 0.20 0.24 Sqbm 1.11e-01 1.42e-01 kl_q 3.93e-04 5.57e-04
%8 lr 8 8 bm 8 8 qbm 8 8 qbm1 8 8 ent 0.69 0.69 Sqbm 6.93e-01 6.93e-01 kl_q  NaN  NaN
%88 lr 2 4 bm 2 4 qbm 1 4 qbm1 1 4 ent 0.56 0.63 Sqbm 5.62e-01 7.84e-01 kl_q 3.56e-01 4.07e-01

%th_org=0.7
%104 lr 0 0 bm 0 0 qbm 0 0 qbm1 0 0 ent 0.00 0.06 Sqbm 1.15e-08 4.02e-02 kl_q 9.57e-15 1.36e-03
%56 lr 2 4 bm 1 4 qbm 0 1 qbm1 0 0 ent 0.00 0.29 Sqbm 1.58e-04 3.63e-01 kl_q 7.31e-08 3.15e-02
%8 lr 8 8 bm 2 3 qbm 0 0 qbm1 0 0 ent 0.00 0.00 Sqbm 2.39e-06 1.50e-05 kl_q 6.13e-12 7.06e-11
%88 lr 2 4 bm 1 4 qbm 0 0 qbm1 0 0 ent 0.00 0.48 Sqbm 5.58e-06 5.10e-01 kl_q 4.82e-11 2.47e-02
%

\begin{table}
\bc
{\small 
\begin{tabular}{c||c|c|c|c||c|c|c|c}
& \multicolumn{4}{c||}{$\theta=0$} & \multicolumn{4}{c}{$\theta=0.7$}\\
& BM & QBM & ENT & KL & BM & QBM & ENT & KL\\\hline
104 &  0	& 0 		& $0.00-0.01$ & $<\num{8e-6}$ & 0 		& 0	& $0.00-0.06$ & $<\num{1e-3}$\\
56  &  2-4 & 0		& $0.20-0.24$ & $<\num{6e-4}$ & 1-4 & 	0 	& $0.00-0.29$ & $<\num{3e-2}$\\
88  	& 2-4 & 1-4	& $0.56-0.63$ & $<\num{4e-1}$ &1-3 & 	0 	& $0.00-0.48$ & $<\num{2e-2}$ \\
8  & 8  & 8 		&$0.69-0.69$ & NA & 2-6 & 	0  &$0.00-0.00$ & $<\num{7e-11}$ 
\end{tabular}
}
\ec
\caption{Classification performance of various methods on all 256 classification problems on 3 binary inputs. 
Problems are partitioned in 4 subsets. Columns BM and QBM lists the range of classification errors (in the range 0 to 8) for that method for that subset of problems. 
ENT lists the range of entropies of the QBM solution $\rho_\mathrm{qbm}$. KL lists the range of KL divergences $KL(q|p)$ with $p(s)=\frac{1}{Z}|\psi(s)|^2$ and $\psi(s)$ the ground state of $H$. 
}
\label{classification}
\end{table}

%\section{$I=2I_c$ when $B$ is determined by $A$}
%\label{IisIc}
%
%Suppose $A$ has states labeled by $x$ and $B$ has states labeled by $y$. Here we show that $I=2I_c$ when $y$ depends deterministically on $x$. 
%Define the probability distribution on $x,y$ as $q(x,y)=q(x)q(y|x)$. Since $y$ depends deterministically on $x$ there exists a function $f$ such that $q(y|x)= \delta_{y,f(x)}$. 
%Classically we have
%\beaa
%h_A^c&=&-\sum_x q(x)\log q(x) \qquad h_B^c=-\sum_y q(y)\log q(y)\\
%h^c_{B|A}&=&0\\
%h^c_{AB}&=& -\sum_{x,y}q(x,y)\log q(x,y)=-\sum_x q(x)\log q(x)=h_A^c\\
%I_c&=&h_A^c+h_B^c-h_{AB}^c=h_B^c
%\eeaa
%with $q(y)=\sum_x q(x)\delta_{y,f(x)}$. In general $h_{AB}^c\ge h_B^c$. This implies that $h_A^c\ge h_B^c$ and $I_c\le \mathrm{min}(h_A^c,h_B^c)=h_B^c$. Thus, $I_c$ is maximized when $A$ determines $B$. 
%
%In the quantum case we have that the marginal density matrix of system $B$ is diagonal: $\eta_B(y,y')=\sum_x \sqrt{q(x,y)q(x,y')}=\delta_{y,y'} q(y)$. Thus, $h_B=h_A=-\sum_y q(y)\log q(y)$ and 
%\beaa
%I=2h_B=2h_B^c=2 I_c
%\eeaa
%NB: $\eta_A$ is not diagonal, unless the relation between $x$ and $y$ is one-to-one. 
%
%

\section{Projective measurements}
\label{projective}
 A projective measurement can be written as a sum $\sum_k \lambda_k E_k$,  with $E_k=\ket{\phi_k}\bra{\phi_k}$  is a set of Hermitian orthogonal projective operators on a Hilbert space that sum to the identity operator: $\sum_k E_k =I$. 
The outcome of the measurement is any of its eigenvalues $\lambda_k$. The measurement outcome is a stochastic event: repeated measurements $E_k$ on the same quantum system with density matrix $\rho$ may yield different values. 
The probability 
\beaa
p(\text{outcome of measurement is}\ \lambda_k)= \Tr \left(E_k\rho\right)\qquad \sum_k \Tr \left(E_k\rho\right)=1
\eeaa

In Eq.~\ref{eta_classical} we defined $\ket{\psi}$ and $\eta$ in the $\sigma^z$ basis $\ket{s}$ that we considered throughout the paper. 
%For the classical data distribution $q(s)$ we defined $\psi(s)=\sqrt{q(s)}$ and the state of the quantum system in Eq.~\ref{eta_classical} as 
%\beaa
%\ket{\psi}=\sum_s \psi(s) \ket{s}\qquad 
%\eta= \ket{\psi}\bra{\psi}=\sum_{s,s'}\eta(s's)\ket{s'}\bra{s}
%\eeaa
%with $\eta(s',s)=\sqrt{q(s')q(s)}$ the components of $\eta$ in the $\sigma^z$ basis $\ket{s}$ that we considered throughout the paper. 
This basis also defines a set of projective measurements 
$E_s=\ket{s}\bra{s}$.
The probability of that measurement $E_s$ on this quantum system yields outcome $1$ is $\Tr\left(E_s \eta\right)=q(s)$. 

The probability of the measurement outcome depends on the choice of measurement basis. 
Another valid set of projective measurement is $E_t=\ket{t}\bra{t}$ with $\ket{t}$ a complete orthogonal basis. 
The probability of outcome $t$ is on the same quantum system $\eta$ is
\bea
%\tilde{q}(t)= \av{t|\psi}\av{\psi|t} \qquad \av{t|\psi}=\sum_s \psi(s) \av{t|s}\label{basistransform}
\tilde{q}(t)=|\tilde{\psi}(t)|^2  \qquad \tilde{\psi}(t) =\sum_s \psi(s) U(t,s)\label{basistransform}
\eea 
and $\sum_t \tilde{q}(t)=1$. $U(t,s)=\av{t|s}$ is a unitary matrix and defines the change of coordinates of $\ket{\psi}$ in the two bases. 
Define $\tilde{\eta}=U \eta U^\dagger$ as the components of $\eta$ in the new basis. Then $\tilde{q}(t)=\tilde{\eta}(t,t)$. 
For any $U$ we get a different set of states $\ket{t}$ and a different classical probability distribution $\tilde{q}(t)$. For $U=I$ we get $\tilde{q}=q$.

\section{Mutual information}
\label{MI}
Here we review classical and quantum mutual information for bipartite systems. 

Suppose that the system of interest is described by variables $s=(s_1,\ldots,s_n)$. We model the interactions between these variables by a probability distribution $q(s)$. We know $q$, we do not know $s$ and
the classical (Shannon) entropy $h_c(q)=-\sum_s q(s)\log q(s)$ quantifies the uncertainty in $s$ given $q$. 
%When $q$ is a very peaked distribution the uncertainty is low and when $q$ is a broad distribution the uncertainty is high. 

Partition the set of variables $s_1,\ldots, s_n$ into two sub sets $A$ and $B$, with $B$ the complement of $A$ and write $s=(s_A,s_B)$ with $s_A$ and $s_B$ the vector of variables in $A$ and $B$, respectively. 
The uncertainty in $s_A$ is given by the entropy $h_c(q_A)$ with $q_A(s_A)=\sum_{s_B} q(s_A,s_B)$ and similar for $s_B$. 

When the sub systems $A$ and $B$ are correlated, observing the state $s_A$ gives us information on $s_B$. The uncertainty in $s_B$ is given by the entropy of the conditional distribution $q(s_B|s_A)$ and depends on the observed value of $s_A$. The conditional entropy is defined as the remaining uncertainty in $B$ when observing $s_A$, averaged over all values $s_A$:
\beaa
h_c(B|A)=-\sum_{s_A} q(s_A) \sum_{s_B} q(s_B|s_A) \log q(s_B|s_A)=h_c(q)-h_c(q_A)
\eeaa
The remaining uncertainty is less that the original uncertainty in $B$: $h_c(B|A)\le h_c(q_B)$. 
The mutual information between $A$ and $B$ is the difference:
\bea
I_c(q)=h_c(q_B)-h_c(B|A)=h_c(q_A) +h_c(q_B)-h_c(q) =KL(q|q_A q_B) \label{Ic}
\eea
where $KL$ is the KL divergence defined in Eq.~\ref{KL} and $q_A q_B$ is the product of marginal distributions $q_A(s_A)q_B(s_B)$. 
$I_c(q)$ quantifies how much the uncertainty in sub system $B$ is reduced on average by observing $s_A$. 
From the last identity it is clear that this also holds with $A$ and $B$ interchanged. 
$I_c(q)$ satisfies
\bea
0\le I_c(q)\le \min(h_c(q_A),h_c(q_B))\label{bound_Ic}
\eea
%The  upper bound on $I_c_{AB}$ can be easily seen by noting that the conditional entropy
%\beaa
%-\sum_{s_A}p(s_A)\sum_{s_B}p(s_B|s_A)\log p(s_B|s_A)=h_c(p)-h_c(p_A)\ge 0 
%\eeaa
%The lower bound follows because $KL(p,p_Ap_B)\ge 0$.

Consider a density matrix $\rho$ on $n$ variables. The quantum mutual information is defined in analogy with Eq.~\ref{Ic} as
\bea
I(\rho)=S(\rho,\rho_A\otimes \rho_B)=h(\rho_A)+h(\rho_B)-h(\rho) \label{I}
\eea
with $h(\rho)$ the von Neumann entropy of $\rho$ as defined in Eq.~\ref{quantum_entropy}, $S$ the relative entropy defined in
Eq.~\ref{Scross} and $h(\rho_A), h(\rho_B)$ the entropies of the reduced density matrices\footnote{
In components, write $s=(s_A,s_B)$ with $s_{A},s_B$ the variables in $A$ and $B$, respectively. Then $\rho(s,s')=\rho(s_A,s_B,s_A',s_B')$ and 
$\rho_A(s_A,s_A')=\sum_{s_B}\rho(s_A,s_B,s_A',s_B)$.}
\bea
\rho_A=\Tr_B\left(\rho\right)\qquad \rho_B =\Tr_A\left(\rho\right)
\eea
For the quantum mutual information one can derive the bounds
\bea
0\le I(\rho) \le 2 \min(h(\rho_A),h(\rho_B))\label{bound_I}
\eea
The lower bound follows from Klein's inequality $S(\eta,\rho)\ge 0$ which holds for any two density matrices $\eta,\rho$ \cite{carlen2010trace}. 
The upper bound follows from the Araki-Lieb inequality  \cite{araki2002entropy}
\bea
h(\rho)\ge |h(\rho_A)-h(\rho_B)| \label{araki}
\eea

Von Neumann (quantum) entropy and quantum mutual information are counter intuitive from a classical point of view. 
For classical systems, when a system $A$ is coupled to $B$, the entropy of the total system $AB$ cannot decrease:  $h_c(q_A)\le h_c(q)$ with $q_A$ the marginal distribution of $q$ on sub system $A$. The intuitive explanation for this is to equate entropy with uncertainty. Then the uncertainty of the total system is always at least as large as the uncertainty of the sub system. 
For quantum systems this is not true. The inequality Eq.~\ref{araki} allows cases where the entropy $h(\rho_A)> h(\rho)$, with
$\rho_A$ the reduced density matrix of $\rho$ on sub system $A$.
Also, classically,  the mutual information that sub system $A$ has about $B$ is always less than the total information in $B$: $I_c(q)\le h_c(q_B)$ (Eq.~\ref{bound_Ic}). 
For quantum systems, since it is possible that $h(\rho_A)> h(\rho)$,
Eq.~\ref{I} implies that $I(\rho)>h(\rho_B)$, ie. sub system $A$ has more information about $B$ than the total information in sub system $B$. 

\section{Singular value decomposition}
\label{schmidt}
Here we review the Singular value decomposition (SVD), also known as Schmidt decomposition, and show that $h(\rho_A)=h(\rho_B)$ when $\rho$ is a pure state. 
Write $s=(s_A,s_B)$ with $s_{A,B}$ the states in $A$ and $B$, respectively. Write $\psi(s_A,s_B)$ as an $n_A\times n_B$ matrix indexed by $s_A,s_B$ with $n_{A,B}$ the number of states in sub system $A,B$, respectively.
Using the SVD \cite{miszczak2011singular} we can write
\bea
\psi =\sum_{k=1}^d \sqrt{\lambda_k} v_k w_k^\dagger\qquad \psi(s_A,s_B)=\sum_{k=1}^d \sqrt{\lambda_k}\  v_k(s_A) w_k^*(s_B) \label{schmidt1} %\quad \text{or} \quad \psi(s_A,s_B)=\sum_{i=1}^q \sqrt{\lambda_i}\  v_i(s_A) w_i(s_B)
\eea
with $\lambda_k$ positive and $v_k,w_k, k=1,\ldots, d$ complex vectors of dimension $n_{A,B}$, respectively and $d=\min(n_A,n_B)$. When $n_A>d$ additional orthogonal vectors $v_k$ are defined arbitrarily to make the basis of $A$ complete and similar for $B$. Then $v_k^\dagger v_l=w_k^\dagger w_l=\delta_{kl}$. From the normalization of $\psi$ it follows that $\sum_k\lambda_k=1$. The reduced density matrices are 
\beaa
\rho_A=\Tr_B(\rho)=\psi\psi^\dagger=\sum_{k=1}^d \lambda_k v_kv_k^\dagger \qquad \rho_B=\Tr_A(\rho)=\psi^\dagger\psi=\sum_{k=1}^d \lambda_k w_k w_k^\dagger
\eeaa
where $\psi\psi^\dagger$ and $\psi^\dagger\psi$ denote matrix products, summing over the inner index. Thus $\rho_A$ and $\rho_B$ have the same eigenvalues and $h(\rho_A)=h(\rho_B)=-\sum_{k=1}^d \lambda_k \log \lambda_k$.

\section{Classical correlations}
\label{cc}

{\em We first define the classical correlations $C$ and show that
$C=I(\rho)/2$ for a pure state \cite{modi2012classical}. }
A projective measurement $\{E_a=\ket{\psi_a}\bra{\psi_a}\}$ on sub system $A$ transforms the density matrix to 
\beaa
\rho \to \rho' =\sum_a \ket{\psi_a}\bra{ \psi_a}\otimes \av{\psi_a| \rho \psi_a}
\eeaa
$\av{\psi_a| \rho \psi_a}$ is a density matrix on $B$ with components on the $\ket{s}$ basis $
\av{\psi_a|\rho\psi_a}(s_B,s_B')=\sum_{s_A,s_A'}\psi_a(s_A) \rho(s_A,s_B;s_A',s_B')\psi_a(s_A')$. 
$A$ observes outcome $a$ with probability 
\beaa
p_a=\Tr \left(\ket{\psi_a}\bra{\psi_a} \rho\right)=\av{\psi_a|\rho_A \psi_a}
\eeaa 
with $\rho_A=\Tr_B(\rho)$ and the marginal density matrix on system $B$ is 
\beaa
\rho_{B|a}=\frac{\av{\psi_a|\rho\psi_a} }{p_a}
\eeaa
The classical correlation is defined in analogy with the classical mutual information  as the difference  $H(B)-H(B|A)$ (first expression in Eq.~\ref{Ic}), maximized over all possible measurements:
\beaa
C=\max_{\{E_a\}} \left(h(\rho_B)-\sum_a p_a h(\rho_{B|a} )\right)
\eeaa
If $\rho$ is a pure state $\rho=\ket{\psi}\bra{\psi}$, we get  $\av{\psi_a| \rho \psi_a}=\ket{\tilde{\psi}_a}\bra{\tilde{\psi}_a}$  with $\tilde{\psi}_a(s_B)=\sum_{s_A}\psi_a(s_A)\psi(s_A,s_B)$. Thus, 
$\rho_{B|a}$ is a pure state and $h(\rho_{B|a})=0$ and $C=h(\rho_B)$. Since $\rho$ is a pure state, $I(\rho)=2h(\rho_B)$ (see Eq.~\ref{Iq}). Therefore $C=I(\rho)/2$. 

The classical mutual information $I_c(\tilde{q})$ depends on the choice of measurement basis. 
{\em We now show that $I_c(\tilde{q})=C$ in the basis where $\tilde{q}$ is diagonal, with $\tilde{q}$ the transformed probability distribution $\tilde{q}$ Eq.~\ref{basistransform}. }
Consider the Schmidt decomposition Eq.~\ref{schmidt1} of $\psi(s_A,s_B)$. 
We identify the Schmidt components $k$ with the states $t_A,t_B$ and define
\beaa
U_A(t_A,s_A)=v_{t_A}(s_A) \qquad U_B(t_B,s_B)=w_{t_B}(s_B)
\eeaa
Since $U_A(t_A,s_A)=\av{t_A|s_A}$ and similar for $B$, these define the coordinates of $\ket{t_A}$ and $\ket{t_B}$ on the original $\ket{s}$ basis. 
With this choice we get from Eq.~\ref{basistransform} that 
\bea
\tilde{\psi}(t_A,t_B)=\sqrt{\lambda_{t_A}}\delta_{t_A,t_B}\qquad \tilde{q}(t_A,t_B)=\lambda_{t_A}\delta_{t_A,t_B}\label{q_svd}
\eea
The basis transformation makes $\tilde{q}$ diagonal.  
The classical mutual information is
\beaa
h_c(\tilde{q})=h_c(\tilde{q}_A)=h_c(\tilde{q}_B)=I_c(\tilde{q})=-\sum_k \lambda_k \log \lambda_k=h(\eta_A)=C
\eeaa
which completes the proof.

{\em We show that $I_c(q)=I(\eta)/2$ when the state of sub system $B$ is determined by the state of $A$ or vice versa.}  When $B$ is determined by $A$, $h_c(B|A)=0$ and 
the classical mutual information is $I_c(q)=h_c(q_B)$. 
Because $s_B$ depends deterministically on $s_A$ we have $q(s_A,s_B)q(s_A,s_B')=q^2(s_A,s_B)\delta_{s_B,s_B'}$. Therefore, the reduced density matrix $\eta_B$ is diagonal:
$\eta_B(s_B,s_B')=\sum_{s_A}\sqrt{q(s_A,s_B)q(s_A,s_B')}e^{i\alpha(s_A,s_B)-i\alpha(s_A,s_B')}=\delta_{s_B,s_B'}q_B(s_B)$ with $q_B(s_B)=\sum_{s_A}q(s_A,s_B)$. Therefore, $h(\eta_B)=h_c(q_B)$ and  $I_c(q)=I(\eta)/2$. 
%Note, that $q$ is not necessarily equal to its SVD decomposition, but $I_c(q)=I_c(\tilde{q})$.
%(see~\ref{examples} where we compute the SVD for the parity problem).

As an example consider that sub system $B$ is a single spin $i$. Then $\eta_B$ can be written in terms of its spin statistics as
\beaa
\eta_B= \frac{1}{2}\left(\begin{tabular}{cc}$1+m_i^z $& $m_i^x-i m_i^y$\\$m_i^x+im_i^y$& $1-m_i^z$\end{tabular}\right)
\eeaa
with $m_i^{x,y,z}=\av{\sigma_i^{x,y,z}}_\eta$. If spin $i$ depends deterministically on (a subset of) the other spins, $\eta_B$ is diagonal and thus $m_i^x=m_i^y=0$. 
%Conversely, we see from Eq.~\ref{spin_statistics} 
%\beaa
%m_i^x=\av{\sigma_i^x}_\eta=2\sum_{s_{\setminus i}} q(s_{\setminus i})\sqrt{q(s_i|s_{\setminus i})q(-s_i|s_{\setminus i})}
%\eeaa
%that if $0\le m_i^x \le 1$. $m_i^x$ is a sum of non-negative terms, and $m_i^x=0$ it implies that $s_i$ is a deterministic function of $s_{\setminus i}$ (either $q(s_i=1|s_{\setminus i})=0$ or $q(s_i=-1|s_{\setminus i})=0$ for each $q(s_{\setminus i})>0$).
This occurs for the parity problem in Figure~\ref{parity_bm} for all $i$, because the value of each spin is fixed once the value of the other spins are given.

\section{The non-locality of quantum physics}
\label{non-local}
\cite{wigner1970hidden} gives a particularly clear explanation of the notion of locality in terms of hidden variables.
Consider a physical system that is composed of sub systems. For classical systems, it is natural to describe the state of the system in terms of local variables that describe the
state of each of the sub systems.  Any composite system that is  described by local variables satisfies the so-called Bell inequalities \cite{bell1964einstein,bell2001problem}. 
Composite quantum systems can violate these inequalities and thus violate the intuitive notion of locality. 

%For mixed states this is no longer true. See for instance \cite{vedral2017foundations}. 

%\cite{horodecki2009quantum} page 873 shows example of bell states for two spin system. 
%the example in this section is called is the ÒcanonicalÓ maximally entangled state. 
%Moreover, Braunstein et al. showed that the Bell states
%are eigenstates of the Bell operator 16 and they maximally violate the Bell-CHSH inequality 17 see Sec.
%IV Braunstein et al., 1992.

%For the data  state $\eta$, the conditional entropy $h(\eta)-h(\eta_A)=-h(\eta_A)$ is negative. This is a quantum mechanical effect that cannot occur in a classical system. 
%The negative entropy is related to the violation of the Bell inequalities \cite{cerf1997entropic}.

Suppose we do $K$ measurements $M_k, k=1,\ldots K$ on a quantum system. The hidden variable idea assumes the existence of additional degrees of freedom that cannot be measured directly, but that determine the measurement outcomes $m_k$. 
Repeating the same measurement on the same quantum system may yield different outcomes. Therefore $m_k$ can be treated as a stochastic variable. The outcomes of different measurements on the same quantum system are correlated and the statistical relation between these outcomes can always be described by a joint probability distribution $p(m_1,\ldots,m_K)$. 

Now, suppose that the system is composed of two sub systems $A$ and $B$.
We can make simultaneous measurements $M_{ij}=A_i\otimes B_j$ on the two sub systems, with $A_i, i=1,\ldots, n_A$ and $B_j, j=1,\ldots, n_B$ the possible measurements on systems $A$ and $B$, respectively.  
Suppose that each measurement $A_i$ and $B_j$ has two possible outcomes. Then each measurement $M_{ij}$ has 4 possible outcomes that are determined by the hidden variable $m_{ij}=(a_{ij},b_{ij})$ with $a_{ij}$ and $b_{ij}$ binary values: for instance measuring $A_1\otimes B_1$ yields possible outcomes $m_{11}=(+,+),(+,-),(-,+)$ or $(-,-)$. The correlations between all possible measurement outcomes can be fully captured by a joint probability distribution
\beaa
p(a_{11},b_{11},a_{12},b_{12},\ldots, a_{n_An_B},b_{n_An_B})
\eeaa
on $2K$ binary variables with $K=n_An_B$. 

The hidden variable $a_{ij}$ defines the outcome of the measurement $A_i$ on sub system $A$ when at the same time a measurement $B_j$ on sub system $B$ is made. 
$a_{ij}$ should obviously depend on the measurement $A_i$, but it is not so clear whether it should depend on the measurement $B_j$, in particular when the systems $A$ and $B$ 
are spatially far removed from each other. 
The Bell assumption of locality is that the measurement outcome $a_{ij}$ does not depend on the measurement that is performed on system $B$: $a_{ij}=a_i$ and similar $b_{ij}=b_j$. This reduces the $2n_An_B$ binary variables to $n_A+n_B$ binary variables and 
\beaa
p(a_{11},b_{11},a_{12},b_{12},\ldots, a_{n_An_B},b_{n_An_B}) \to p(a_1,\ldots, a_{n_A},b_1,\ldots, b_{n_B})
\eeaa
The locality assumption implies that the outcomes of quantum measurement can be described by a probability distribution on local variables. The construction by Bell \cite{bell1964einstein} 
shows that the correlations produced by measurements on a quantum system can violate this assumption. The counter intuitive conclusion is that the statistics of outcomes of measurement $A_i$ on system $A$ depends on what measurement is performed system $B$. In other words, the correct statistical description requires the non local variables $a_{ij}, b_{ij}$.

\clearpage

\end{document}